\documentclass[aps, prl, twocolumn, groupedaddress]{revtex4-1}
\usepackage{graphicx}
\usepackage{braket}
\usepackage{amsmath}
\usepackage{dcolumn}
\newcommand*\us{\mskip3mu}
\newcommand*{\tx}[1]{\mathrm{#1}}

\begin{document}
\title{Quantum interface between photonic and superconducting qubits}

\author{Yuta Tsuchimoto, Patrick Kn\"{u}ppel, Aymeric Delteil, Zhe Sun, Martin Kroner,}
\author{Ata\c{c} Imamo\u{g}lu}
\affiliation{Institute of Quantum Electronics, ETH Zurich, CH-8093 Z\"{u}rich, Switzerland}

\date{\today}

\begin{abstract}
We show that optically active coupled quantum dots embedded in a superconducting microwave cavity can be used to realize a fast quantum interface between photonic and transmon qubits. Single photon absorption by a coupled quantum dot results in generation of a large electric dipole, which in turn ensures efficient coupling to the microwave cavity. Using cavity parameters achieved in prior experiments, we estimate that bi-directional microwave-optics conversion in nanosecond timescales with efficiencies approaching unity is experimentally feasible with current technology. We also outline a protocol for in-principle deterministic quantum state transfer from a time-bin photonic qubit to a transmon qubit. Recent advances in quantum dot based quantum photonics technologies indicate that the scheme we propose could play a central role in connecting quantum nodes incorporating cavity-coupled superconducting qubits.
\end{abstract}


\maketitle

\paragraph{\label{sc:introduction}Introduction.}
A quantum interface between flying photonic and stationary matter qubits is widely regarded as an essential element of quantum networks~\cite{kimble2008quantum, cirac1997quantum, cirac1999distributed}. Remarkable advances over the last decade have established that circuit-QED, consisting of superconducting (SC) qubits non-perturbatively coupled to a common microwave (MW) cavity, is particularly promising for realization of small-scale quantum information processors~\cite{nakamura1999coherent, devoret2013superconducting}. The most prominent limitation in realization of quantum networks consisting of circuit-QED based processors is the difficulty in transferring quantum information over distances exceeding meters. Motivated by overcoming this roadblock, several groups have embarked on research aimed at creating a quantum interface between SC qubits and propagating photonic qubits. Among the several ingenuous proposals \cite{strekalov2009microwave, rueda2016efficient,hafezi2012atomic, petrosyan2009reversible,williamson2014magneto, fernandez2015coherent,imamouglu2009cavity, blum2015interfacing,hisatomi2016bidirectional,das2017interfacing} to resolve this conundrum, the approach based on using optomechanical coupling \cite{tian2015optoelectromechanical, bochmann2013nanomechanical, balram2016coherent, bagci2014optical, andrews2014bidirectional} has proven to be particularly successful: pioneering experiments have demonstrated conversion efficiency of $10 \%$ with a bandwidth of $30\us\text{kHz}$ \cite{andrews2014bidirectional}. A limitation for most if not all of these approaches is the relatively small effective coupling strength between the single optical and MW photons, which in turn prevents conversion of quantum information on time-scales much shorter than typical SC qubit coherence times.

In this Letter, we propose a novel quantum interface consisting of a coupled quantum dot (CQD) embedded in a low-Q optical cavity and positioned at the antinode of a high-Q SC MW resonator. Unlike the aforementioned hybrid quantum systems, CQDs in integrated structures such as the one depicted in Fig.~1 have large coupling strengths to both optical and MW fields, which ensures fast and high-efficiency bi-directional MW-to-optics conversion. Three key features of the scheme we detail are (i) the use of low-Q asymmetric optical cavity to allow for high-efficiency absorption of an incoming single-photon pulse; (ii) the creation of a large electric dipole in the CQD by absorption of a single photon ensuring strong coupling to the MW resonator; and (iii) fast radiative decay rate of the CQD optical transitions yielding nanosecond time-scale inter-conversion. After presenting a detailed analysis of the quantum interface between MW and optical photons, we describe a scheme for quantum state transfer from an incident flying photon qubit to a SC transmon qubit via the MW resonator.

\paragraph{Structure of the interface and coupling between microwave and optical photons. }
\begin{figure}[ht]
  \begin{center}

    \includegraphics[width=\columnwidth]{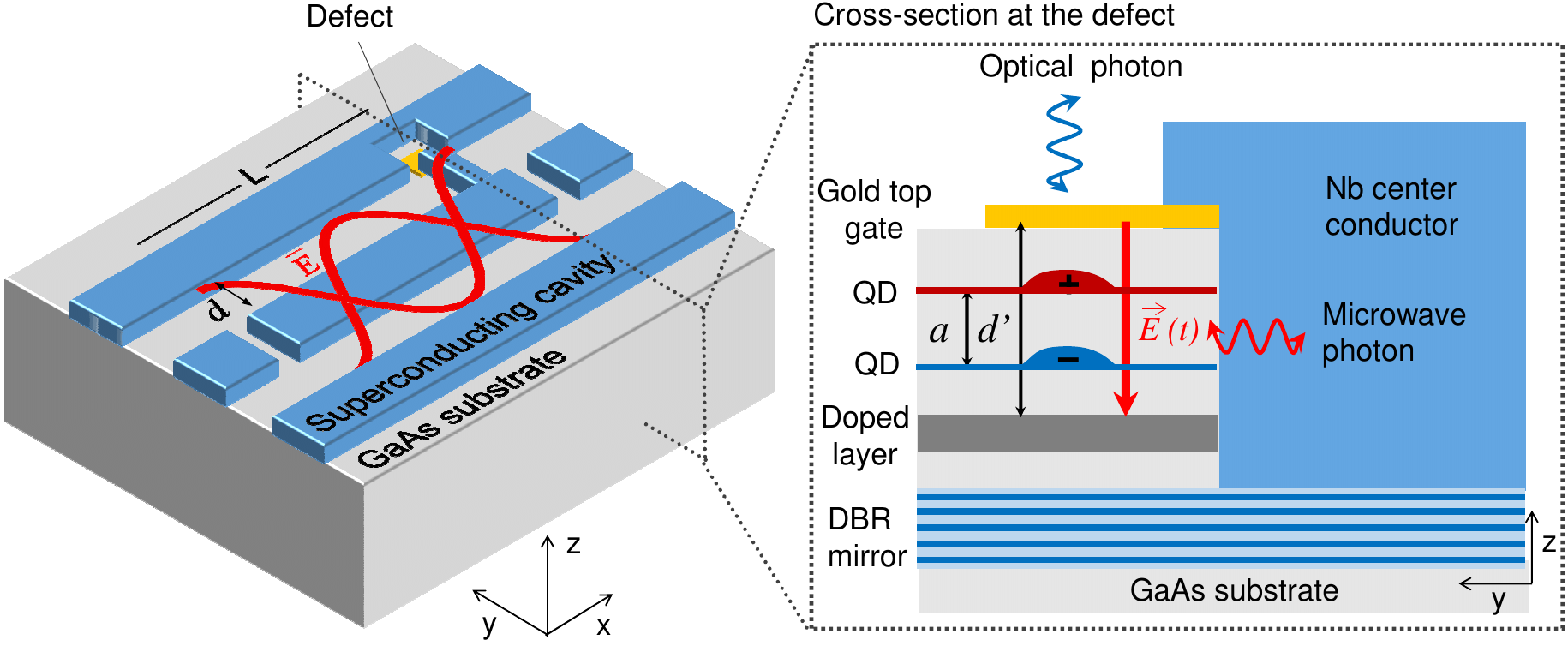}

  \end{center}
  \caption{Schematic of the proposed device consisting of a superconducting (SC) MW cavity incorporating  coupled quantum dots (CQD). In contrast to the standard devices, the SC cavity has a $\mu$m sized defect region at the field antinode that allows for efficient coupling to one of the CQDs. The cross-section image shows the defect geometry cut along the black dashed line: the vertically stacked blue and red  dots form a CQD. An applied DC voltage controls coherent electron tunneling between the two dots. The bottom Distributed Bragg Reflector (DBR) and the top gold layer act as a low quality factor optical cavity that allows for efficient in-and-out coupling of optical photons.}
  \label{F1}
\end{figure}

Figure~\ref{F1} shows the structure that we analyze: the substrate of the SC cavity is an MBE grown GaAs sample consisting of a Distributed Bragg Reflector (DBR) mirror, an n++-GaAs section and a layer of InGaAs CQDs. Before fabrication of the SC coplanar resonator by Nb deposition, the top section of the GaAs substrate containing the n++-GaAs section and the CQDs is etched away to reduce MW losses. To ensure efficient CQD-MW coupling while minimizing MW losses, a defect region of width $\sim 1 \mu$m is introduced at one of the resonator antinodes (Fig.~\ref{F1}): it is only in this small area region of the device that the layers containing CQDs and the n++-GaAs section are not etched away. The top gold layer in the defect, connected to the Nb center conductor, acts as a top gate for adjusting the CQD energy levels to ensure optimal optical coupling and dipole generation. Concurrently, this top $\sim 20$~nm gold layer, together with the bottom DBR mirror forms a low-Q optical cavity that ensures efficient interface between the CQD and the single-photon pulses~\cite{Delley2017}. 

We remark that the device depicted in Fig.~\ref{F1} is motivated by the structure successfully used to demonstrate strong coupling between an electrically defined GaAs CQD charge qubit and an Al coplanar resonator~\cite{Frey2012,Stockklauser2017}. In these experiments, MW resonator Q-factors of several thousands were demonstrated in structures with a defect region incorporating a two-dimensional electron system (2DES). In parallel, large single-photon absorption induced electric dipoles have been demonstrated using CQDs that allow for electric field control of coherent resonant tunneling between the two QDs comprising the CQD~\cite{krenner2005}. As a result, the use of high impedance SC cavity with vacuum-field enhancement at the location of the CQD would enable efficient coupling of the large CQD dipole to the SC cavity (up to $\sim 200$~MHz, see Supplemental Material). Finally, low-Q optical cavities with a leaky top mirror have been used in experiments demonstrating distant quantum dot (QD) spin entanglement~\cite{delteil2016} as well as absorption of a photonic qubit by a single QD~\cite{delteil2017}.  The low-Q cavity modifies the radiation pattern of the CQD and provides an excellent match to the Gaussian profile of the incident photon mode, thereby ensuring efficient absorption of the optical photons.


\paragraph{\label{sc:Single photon conversion}Single photon conversion.}
\begin{figure}
\includegraphics[width=\columnwidth]{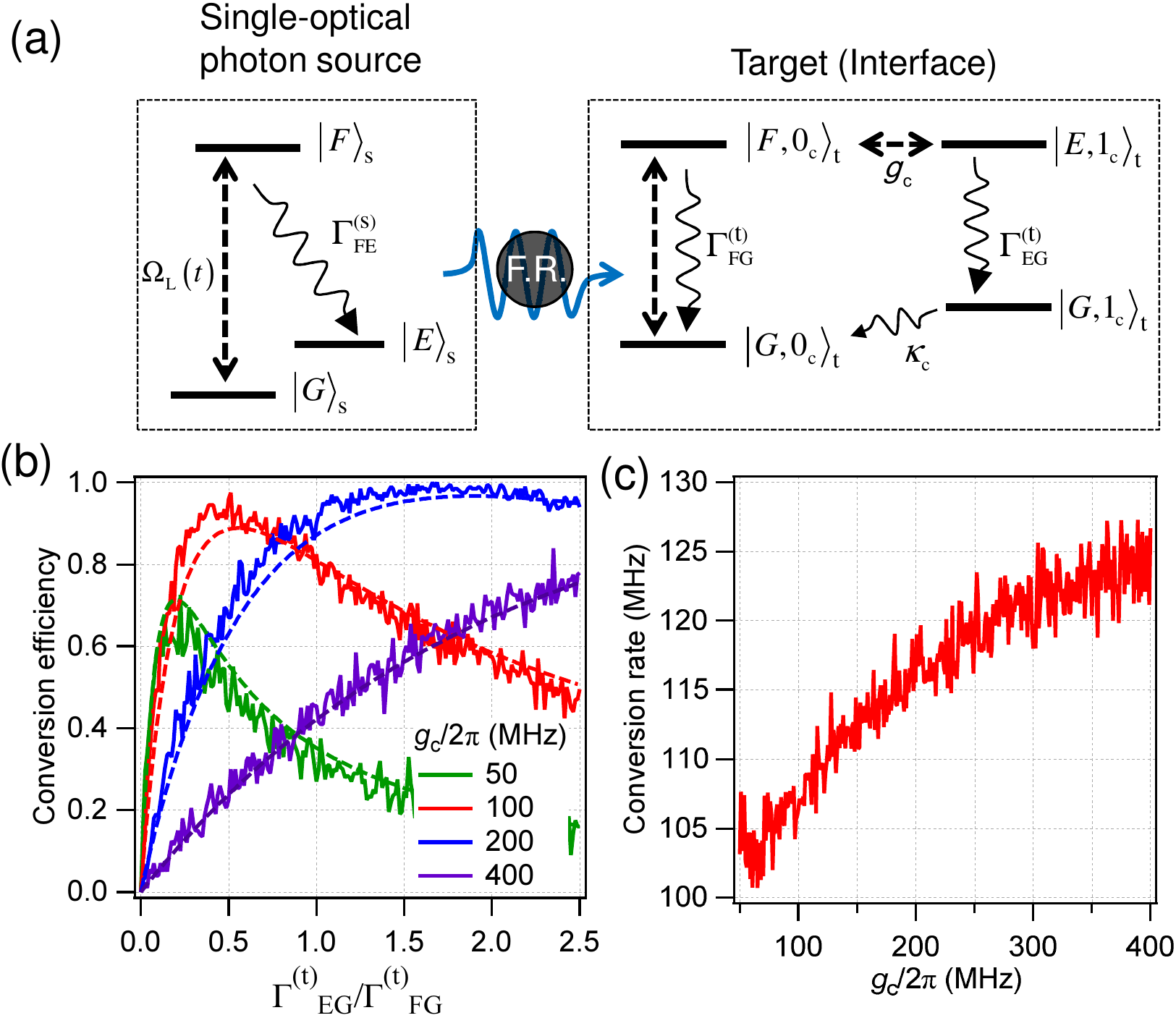}
\caption{\label{F2}(a) Schematic of optical to MW photon conversion. A single photon emitter (left) generates an optical-photon pulse upon coherent drive with laser Rabi frequency $\Omega_\tx{L}$. The generated photon resonantly couples to the target (right) consisting of a CQD and A SC cavity. F.R. indicates a Faraday rotator ensuring unidirectional coupling. An absorbed optical photon at the target is converted into a MW photon in the SC cavity mode through the coupling $g_{\tx{c}}$ between the optically induced dipole and the MW cavity. (b) Conversion efficiency as a function of the decay rates $\Gamma_\tx{EG}^{(\tx{t})}/\Gamma_\tx{FG}^{(\tx{t})}$ for different coupling strengths $g_{\tx{c}}$. The solid lines show simulated results using the quantum Monte Carlo method. Here we set $\Gamma_\tx{FG}^{(\tx{t})}/2\pi = 300\us\tx{MHz}$ and a SC cavity decay rate $\kappa_\tx{c}/2\pi = 3\us\tx{MHz}$. The dashed lines are analytical results under the assumption where weak coherent light is incident on the target. (c) Conversion rate as a function of $g_\tx{c}$ at $\Gamma_\tx{EG}^{(\tx{t})}/\Gamma_\tx{FG}^{(\tx{t})}=1.0$.}
\end{figure}

To analyze conversion of a single optical photon pulse into a MW cavity photon, we consider a QD single photon source (labeled $s$) whose output is channeled to a target CQD (labeled $t$) that is embedded under the defect region of the SC resonator. Figure~\ref{F2}(a) shows the energy level diagram of the source QD (left) and the target CQD (right). A single photon pulse is generated on the $\ket{F}_\tx{s}$ to $\ket{E}_\tx{s}$ transition: the laser pulse with Rabi frequency $\Omega_\tx{L}$ exciting the QD from $\ket{G}_\tx{s}$ to $\ket{F}_\tx{s}$ determines the pulse shape. The transition energies are adjusted using electric and magnetic fields to ensure that the center frequency of the single-photon pulse matches the $\ket{G}_\tx{t}$ to $\ket{F}_\tx{t}$ transition of the target CQD~\cite{delteil2017}. To simplify the analysis, we assume that either a Faraday rotator or a chiral waveguide is placed in between the source and target QDs; this assumption allows us to use the cascaded quantum systems formalism to calculate the conversion efficiency in the limit of low-loss photon transfer~\cite{carmichael1993quantum,Gardiner1993,pinotsi2008single}.

We assume that the target CQD is neutral and its energy levels are tuned using an external gate voltage to ensure coherent tunnel coupling between the lowest electronic states of the top and bottom QDs. The resulting symmetric and antisymmetric excitonic states are labeled $\ket{E}_\tx{t}$ and $\ket{F}_\tx{t}$, respectively. Furthermore, we assume that $\ket{E}_\tx{t}$-$\ket{F}_\tx{t}$ splitting matches the SC resonator resonance frequency. The sequence for optics-to-MW conversion is then \[ \ket{G,0_\tx{c},1_\tx{s},0_\tx{t}}_\tx{t} \rightarrow \ket{F,0_\tx{c},0_\tx{s},0_\tx{t}}_\tx{t} \rightarrow \ket{E,1_\tx{c},0_\tx{s},0_\tx{t}}_\tx{t} \rightarrow \ket{G,1_\tx{c},0_\tx{s},1_\tx{t}}_\tx{t} \] where $\ket{1_\tx{c}}$ denotes the single-MW-photon eigenstate of the SC resonator. $\ket{1_\tx{s}}$ and $\ket{1_\tx{t}}$ denote the single-photon pulses generated by the source QD and the target CQD, respectively.

We calculate the efficiency and speed of the conversion process, determined by the CQD-resonator coupling strength as well as the spontaneous emission rates, using the quantum Monte Carlo method \cite{carmichael1993quantum, pinotsi2008single} (see the supplemental material for a detailed description of the calculation). Since generation of a target photon $\ket{1_\tx{t}}$ upon spontaneous emission on the $\ket{E,1_\tx{c}}_\tx{t} \rightarrow \ket{G,1_\tx{c}}_\tx{t}$ transition heralds successful optical-MW photon conversion, we determine the number of these photon emission events in quantum trajectory simulations to estimate the conversion efficiency and rate.

Figure~\ref{F2}(b) shows the conversion efficiency (solid curves) as a function of a ratio $\Gamma_\tx{EG}^{(\tx{t})}/\Gamma_\tx{FG}^{(\tx{t})}$ for various $g_\tx{c}$.  The efficiency rapidly increases as $\Gamma_\tx{EG}^{(\tx{t})}/\Gamma_\tx{FG}^{(\tx{t})}$ increases from 0 and reaches a maximum at a certain $\Gamma_\tx{EG}^{(\tx{t})}/\Gamma_\tx{FG}^{(\tx{t})}$. We attribute this dependence to a quantum interference between the incident single-photon pulse and the secondary field generated at the same frequency by the target CQD (see the Supplemental Material). When photon detection at the $\ket{F,0_\tx{c}}_\tx{t} \rightarrow \ket{G,0_\tx{c}}_\tx{t}$ transition is suppressed due to this destructive interference, the efficiency reaches its maximum value. One can see that the maximum shifts to larger $\Gamma_\tx{EG}^{(\tx{t})}/\Gamma_\tx{FG}^{(\tx{t})}$ ratios and approaches unity as we increase $g_\tx{c}$.

To obtain an analytical expression valid in the limit of purely coherent light scattering (see Supplemental Material), we note that the conversion efficiency $\zeta$ can be expressed as
\begin{equation}
\zeta=1-\braket{b_\tx{out}^\tx{n\dagger} b_\tx{out}^\tx{n}},
\end{equation}
where $\braket{b_\tx{out}^\tx{n}}$ is the normalized mean field generated at the $\ket{F,0_\tx{c}}_\tx{t} \rightarrow \ket{G,0_\tx{c}}_\tx{t}$ transition and is given by
\begin{equation}
\braket{b_\tx{out}^\tx{n}}=1-\frac{2\Gamma_\tx{EG}^{(\tx{t})}}{4g_\tx{c}^2/\Gamma_\tx{FG}^{(\tx{t})}+\Gamma_\tx{EG}^{(\tx{t})}-ig_\tx{c}},
\label{Eref}
\end{equation}
assuming SC cavity decay rate $\kappa_c$ is much smaller than $g_c$ and all spontaneous emission rates. The dashed lines in Fig. \ref{F2}(b) are the analytical results. Here, $4g_\tx{c}^2/\Gamma_\tx{FG}^{(\tx{t})}$ is the effective decay rate from $\ket{E,1_\tx{c}}_\tx{t}$ to $\ket{G,0_\tx{c}}_\tx{t}$. The real part of Eq. (\ref{Eref}) expresses the quantum interference. If the two decay rates $4g_\tx{c}^2/\Gamma_\tx{FG}^{(\tx{t})}$ and $\Gamma_\tx{EG}^{(\tx{t})}$ are equal, the reflection is suppressed, i.e. the efficiency is maximal. The imaginary part $ig_\tx{c}$ accounts for the frequency detuning between the incident photon and $\ket{F,0_\tx{c}}_\tx{t}$ which is shifted by $g_\tx{c}$ due to the MW coupling. This detuning limits the maximum efficiency when $g_\tx{c} > 4g_\tx{c}^2/\Gamma_\tx{FG}^{(\tx{t})}$ as shown in Fig.~\ref{F2}(b). In practice, it is possible to restore the efficiency in the limit of large $g_\tx{c}$ to almost unity by tuning the frequency of the incident photon. The condition for the efficient conversion is, therefore, $g_\tx{c}^2 = \Gamma_\tx{FG}^{(\tx{t})}\Gamma_\tx{EG}^{(\tx{t})}/4$. For $\Gamma_\tx{EG}^{(\tx{t})}/\Gamma_\tx{FG}^{(\tx{t})} \sim 1$, $g_\tx{c}/2\pi \sim 100 - 200\us\tx{MHz}$ satisfies this condition, and the efficiency approaches unity.

We plot the calculated conversion rate, defined as the reciprocal of the time required to complete the transfer, in Fig. \ref{F2}(c) as a function of $g_\tx{c}$ at $\Gamma_\tx{EG}^{(\tx{t})}/\Gamma_\tx{FG}^{(\tx{t})} = 1.0$. The rate becomes larger as  $g_\tx{c}$ is increased. When $g_\tx{c}/2\pi \sim 100 - 200\us\tx{MHz}$, the conversion rate is about 110 MHz.

\begin{figure}
\includegraphics[width=0.9\columnwidth]{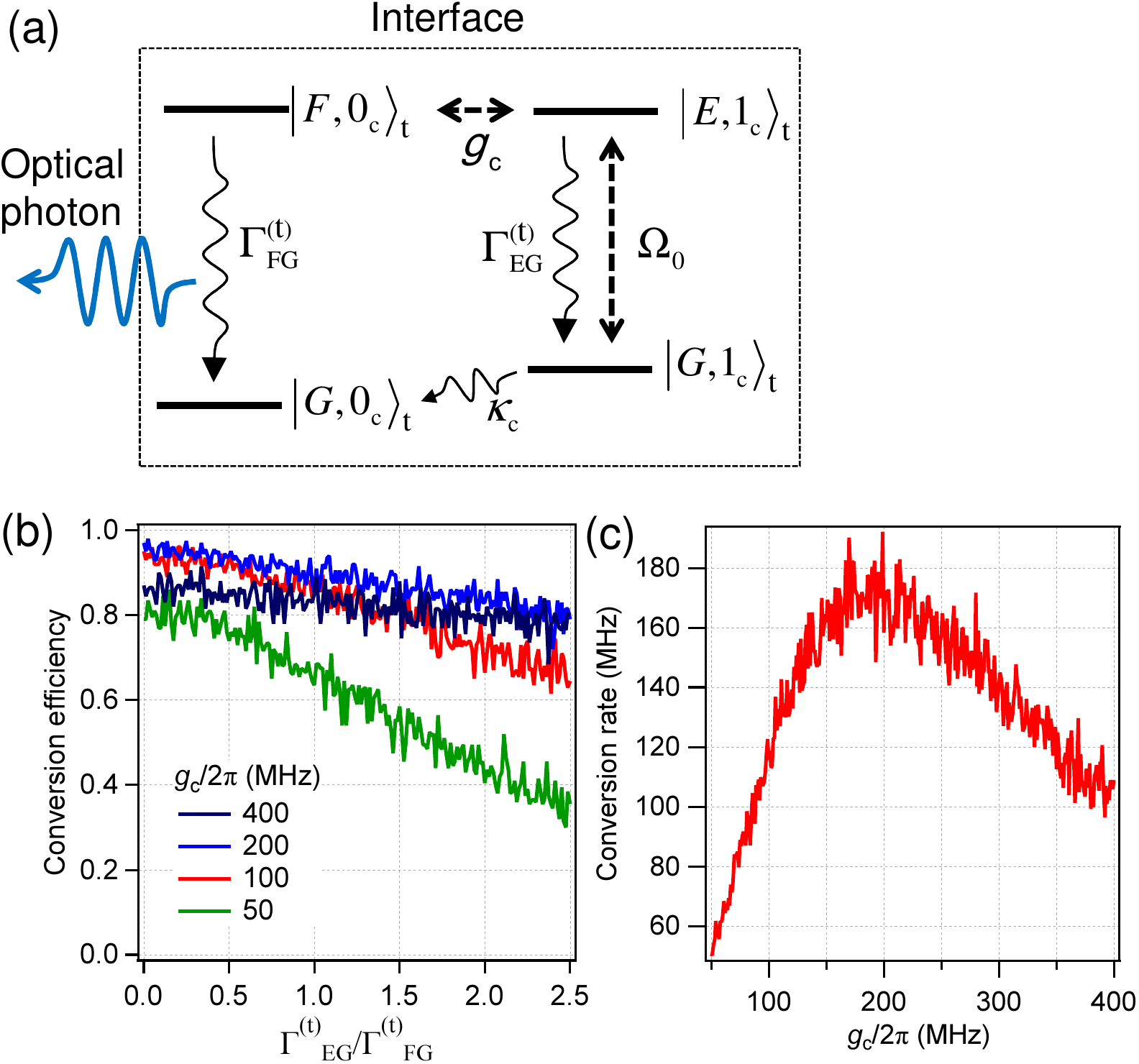}
\caption{\label{F3}(a) Schematic of the MW-to-optical conversion scheme. The conversion starts out in $\ket{G,1_\tx{c}}_{\tx{t}}$. By driving the transition $\ket{G,1_\tx{c}}_{\tx{t}} \rightarrow \ket{F,1_\tx{c}}_{\tx{t}}$ with laser Rabi frequency $\Omega_\tx{0}$, MW photons are upconverted through the cavity-CQD coupling $g_{\tx{c}}$. The converted optical photons are emitted from the $\ket{F,0_\tx{c}}_{\tx{t}} \rightarrow \ket{G,0_\tx{c}}_{\tx{t}}$ transition. (b) Conversion efficiency calculated using the quantum Monte Carlo method as a function of $\Gamma_\tx{EG}^{(\tx{t})}/\Gamma_\tx{FG}^{(\tx{t})}$ for different $g_{\tx{c}}$. We set $\Gamma_\tx{FG}^{(\tx{t})}/2\pi = 300\us\tx{MHz}$ and $\kappa_\tx{c}/2\pi = 3\us\tx{MHz}$. (c) Conversion rate as a function of $g_\tx{c}$ at $\Gamma_\tx{EG}^{(\tx{t})}/\Gamma_\tx{FG}^{(\tx{t})}=1.0$. }
\end{figure}

Next, we consider the reverse process of frequency conversion of single SC resonator MW photon  to a propagating single-photon pulse. Figure \ref{F3}(a) depicts the energy level diagram of the CQD used to realize the quantum interface, which is identical to that of the target CQD in Fig.~\ref{F1}. In contrast to the previous discussion, we now assume that the system starts out in state $\ket{G,1_c}_\tx{t}$ and is excited by a laser field tuned into resonance with the $\ket{G,1_c}_\tx{t} \rightarrow \ket{E,1_c}_\tx{t}$ transition. We emphasize that a fundamental limitation on the conversion efficiency stems from the SC resonator decay rate $\kappa_c$. In the limit where MW-assisted laser up-conversion rate $4g_\tx{c}^2/(\Gamma_\tx{EG}^{(\tx{t})} + \Gamma_\tx{FG}^{(\tx{t})})$ is much larger than $\kappa_c$, conversion efficiency will approach unity. The ratio $\Gamma_\tx{EG}^{(\tx{t})}/\Gamma_\tx{FG}^{(\tx{t})}$ determines the number of photons scattered on the $\ket{G,1_c}_\tx{t} \rightarrow \ket{E,1_c}_\tx{t}$ transition before MW-to-optical photon conversion is successful. If the latter photons are not detected, the indistinguishability of the generated single optical photon pulse will be compromised.  Figure~\ref{F3}(b) shows the dependence of the conversion efficiency as a function of $\Gamma_\tx{EG}^{(\tx{t})}/\Gamma_\tx{FG}^{(\tx{t})}$, determined by counting the number of photon emission events at the $\ket{F,0_c}_\tx{t} \rightarrow \ket{G,0_c}_\tx{t}$ transition in quantum trajectory simulations. For $\Gamma_\tx{EG}^{(\tx{t})}/\Gamma_\tx{FG}^{(\tx{t})} < 0.2$, conversion efficiencies exceeding $90\%$ can be reached even for $\kappa_c/2\pi \sim 3$~MHz. We emphasize that in this limit we expect the generated photons to be highly indistinguishable.

Figure~\ref{F3}(c) shows the conversion rate as a function of $g_\tx{c}$ at $\Gamma_\tx{EG}^{(\tx{t})}/\Gamma_\tx{FG}^{(\tx{t})} = 1.0$. For $g_\tx{c} < \Gamma_\tx{FG}^{(\tx{t})}$, the rate increases with increasing $g_\tx{c}$ and reaches about 170 MHz at $g_\tx{c}/2\pi \sim 200\us\tx{MHz}$. This behavior follows the rate $4g_\tx{c}^2/(\Gamma_\tx{EG}^{(\tx{t})}+\Gamma_\tx{FG}^{(\tx{t})})$. The rate decreases for $g_\tx{c} > \Gamma_\tx{FG}^{(\tx{t})}$ due to the detuning of the laser field from the $\ket{G,1_c}_\tx{t} \rightarrow \ket{E,1_c}_\tx{t}$ transition; as we argued before, this reduction does not represent a real limitation and can be remedied by adjusting the incident laser frequency. For $\Gamma_\tx{EG}^{(\tx{t})}/\Gamma_\tx{FG}^{(\tx{t})} \sim 1.0$, $g_\tx{c}/2\pi \sim 200\us\tx{MHz}$ leads to an efficiency of $\simeq 0.9$ with a conversion rate of 170 MHz.

\paragraph{\label{sc:Quantum state transfer}Quantum state transfer.}
\begin{figure}
\includegraphics[width=\columnwidth]{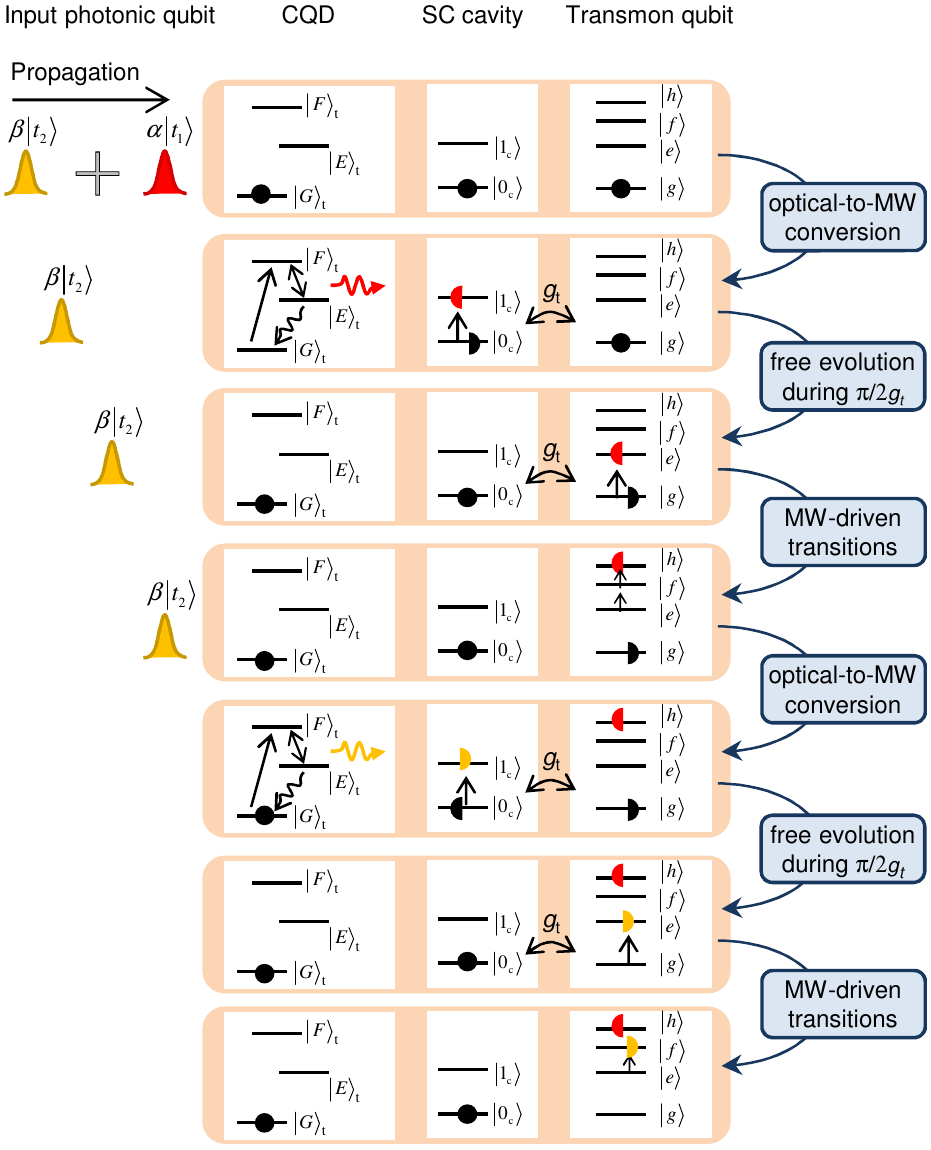}
\caption{\label{F4} Schematic of the proposed scheme for faithful quantum state transfer between photonic time-bin and SC transmon qubits. The system consists of a CQD coupled to a SC cavity, realizing the bi-directional photon conversion, and a transmon qubit coupled to the same SC cavity. The CQD has three energy levels as already shown in Figs. \ref{F2} and \ref{F3}. The transmon qubit has a ground state $\ket{g}$ and higher excited states $\ket{e}$, $\ket{f}$, and $\ket{h}$. The $\ket{F}_\tx{t} \rightarrow \ket{E}_\tx{t}$ transition of the CQD resonantly couples to the SC cavity, which is itself resonant with the fundamental transition of the transmon qubit. Transfer from an optical qubit state $\alpha \ket{t_1} + \beta \ket{t_2}$ to the transmon qubit state $\alpha \ket{h} + \beta \ket{f}$ follows the six steps depicted above, including for each component: an optical-to-MW conversion, free evolution during half the Rabi period, and coherent rotation of the transmon qubit. One can implement the reverse transmon-to-photonic qubit transfer process by a reverse sequence.}
\end{figure}

The bi-directional single photon conversion enabled by the device we propose opens the way for transferring quantum information from a photonic qubit to a SC qubit coupled to the MW cavity, and vice-versa. In the following we propose a protocol realizing such transfer from a time-bin photonic qubit to a transmon qubit. The use of time-bin qubits enables the use of the aforementioned conversion process without modifications, and does not rely on additional local degrees of freedom of the CQD and/or the SC cavity. Commonly used photonic qubits (polarization qubits, dual rail qubits) can be straightforwardly converted into time-bin qubits using linear optics. Finally, while we assume a transmon qubit, any SC qubit with an anharmonic spectrum could be employed.

Figure~\ref{F4}(a) depicts the transfer protocol. The system we consider is almost identical to the device shown in Fig.~\ref{F1}; the only modification is the addition of a transmon qubit coupled to the cavity. We consider the first four levels of the anharmonic qubit spectrum: the ground state $\ket{g}$ and the three first excited states $\ket{e}$, $\ket{f}$ and  $\ket{h}$. The $\ket{g} \rightarrow \ket{e}$ transition is resonant with the cavity mode, with a coupling strength $g_t$ satisfying $\kappa_c \ll g_t < g_c$~\cite{PhysRevA.92.063801}. moreover, we assume that resonant MW pulses can be applied to drive coherent rotations within the three pairs of successive states of the transmon qubit~\cite{peterer2015}. The input photonic state is a time-bin qubit of the form $\alpha \ket{t_1} + \beta \ket{t_2}$, with $t_1$ and $t_2>t_1$ denoting the arrival time of the two components of the qubit. The transmon qubit is initially in $\ket{g}$. Within a few nanoseconds after $t_1$, the $\ket{t_1}$ component is transferred to a MW cavity photon with probability amplitude $\alpha$. The coupled cavity-transmon system will then start to undergo Rabi oscillations at frequency $g_t$. After half a Rabi oscillation period, the MW cavity population is swapped with that of the $\ket{e}$ state of the transmon. At this point, we apply a $\pi$-pulse resonant with the $\ket{e} \rightarrow \ket{f}$ transition followed by a second $\pi$-pulse resonant with the $\ket{f} \rightarrow \ket{h}$ transition, which coherently transfers the population onto the state $\ket{h}$ within a few nanoseconds. Similarly, after $t_2$ the probability amplitude of the second component of the qubit is mapped onto the state $\ket{f}$ using coherent transfer from $\ket{g}$ to $\ket{e}$ followed by a $\pi$-pulse driving the $\ket{e} \rightarrow \ket{f}$ transition. The final state of the transmon is $\alpha\ket{h}+\beta\ket{f}$. In addition, in case of finite transfer efficiency, an unsuccessful transfer process will be associated with finite population of the transmon ground state, whose readout acts as a syndrome measurement for a failed transfer. The total duration of the state transfer can be as short as a few tens of nanoseconds -- much shorter than the coherence time of the transmon qubits (typically several 10s to 100s of microseconds~\cite{koch2007charge, chang2013improved, rigetti2012superconducting, paik2011observation, barends2013coherent}).  In our scheme, population transfer into the $ \left\lbrace \ket{f} , \ket{h} \right\rbrace $ subspace of the transmon allows to decouple the qubit from the MW cavity. The two lower energy states can however be used to store the final state provided that the cavity is tuned away from the $\ket{g} \rightarrow \ket{e}$ transition once the transfer is achieved. An alternative way to transfer the MW cavity population to the transmon, rather than using the excited transmon states, is to make use of several transmon qubits and perform SWAP operation between them to store transferred populations~\cite{dewes2012}.

We remark that an optical-to-MW photon conversion process leads to subsequent emission of a photon from the $\ket{E,1_\tx{c}}_\tx{t} \rightarrow \ket{G,1_\tx{c}}_\tx{t}$ transition of the CQD. Hence after the transfer, the state of the transmon qubit will be entangled with the time-bin degree of freedom of the emitted photon, leading to leakage of which-path information. This qubit-photon entanglement can be used as a resource for generation of entanglement between distant nodes. For the purpose of quantum state transfer, it is possible to erase this entanglement by delaying the first component of the emitted photon by $t_2 - t_1$ and then erase the which-path information by combining the the two paths using a beam-splitter (see Supplemental Material). In case of finite optical losses and/or finite detection efficiencies, such a scheme will not succeed all the time but success will be heralded.

The reciprocal transmon-to-photonic qubit transfer process can be realized using a reverse sequence: the first step in the protocol is the transfer of the amplitudes in $\ket{f}$ and $\ket{h}$ using a sequence of MW $\pi$-pulses applied at $t_1$ and $t_2$, which allows for generation of a finite probability amplitude for a single-MW-photon state due to coherent transmon-cavity interaction. The MW-to-optical photon conversion described earlier is used in the second step to successively upconvert the MW-photon to the optical domain, ensuring faithful state transfer to a time-bin photonic qubit.

In summary, we proposed a device realizing coherent bi-directional photon conversion using CQD coupled with a MW resonator. Simulations based on quantum Monte Carlo method reveal conversion rates up to hundreds of MHz with close-to-unity conversion efficiencies for both up- and down-conversion. The requisite cavity-CQD coupling strengths $g_\tx{c} \sim 200\us\tx{MHz}$ can be achieved by using a high impedance SC cavity~\cite{Stockklauser2017} and enhancing the cavity vacuum electric field at the location of the CQD (see Supplemental Material). By adding a transmon qubit coupled to the same SC cavity, we show that quantum state transfer from photonic qubit to transmon qubit is achievable at rates of several tens of MHz. We believe that such a structure would open the way to high bandwidth quantum networks consisting of SC-qubit-based nodes remotely connected by optical photons.

\begin{acknowledgments}

Y.T., P.K., and A.D. contributed equally to this work. This work is
supported by NCCR Quantum Photonics (NCCR QP), research instrument
of the Swiss National Science Foundation (SNSF), and by by Swiss NSF
under Grant No. 200020-159196.

\end{acknowledgments}

\bibliography{refs}

\begin{thebibliography}{37}%
\makeatletter
\providecommand \@ifxundefined [1]{%
 \@ifx{#1\undefined}
}%
\providecommand \@ifnum [1]{%
 \ifnum #1\expandafter \@firstoftwo
 \else \expandafter \@secondoftwo
 \fi
}%
\providecommand \@ifx [1]{%
 \ifx #1\expandafter \@firstoftwo
 \else \expandafter \@secondoftwo
 \fi
}%
\providecommand \natexlab [1]{#1}%
\providecommand \enquote  [1]{``#1''}%
\providecommand \bibnamefont  [1]{#1}%
\providecommand \bibfnamefont [1]{#1}%
\providecommand \citenamefont [1]{#1}%
\providecommand \href@noop [0]{\@secondoftwo}%
\providecommand \href [0]{\begingroup \@sanitize@url \@href}%
\providecommand \@href[1]{\@@startlink{#1}\@@href}%
\providecommand \@@href[1]{\endgroup#1\@@endlink}%
\providecommand \@sanitize@url [0]{\catcode `\\12\catcode `\$12\catcode
  `\&12\catcode `\#12\catcode `\^12\catcode `\_12\catcode `\%12\relax}%
\providecommand \@@startlink[1]{}%
\providecommand \@@endlink[0]{}%
\providecommand \url  [0]{\begingroup\@sanitize@url \@url }%
\providecommand \@url [1]{\endgroup\@href {#1}{\urlprefix }}%
\providecommand \urlprefix  [0]{URL }%
\providecommand \Eprint [0]{\href }%
\providecommand \doibase [0]{http://dx.doi.org/}%
\providecommand \selectlanguage [0]{\@gobble}%
\providecommand \bibinfo  [0]{\@secondoftwo}%
\providecommand \bibfield  [0]{\@secondoftwo}%
\providecommand \translation [1]{[#1]}%
\providecommand \BibitemOpen [0]{}%
\providecommand \bibitemStop [0]{}%
\providecommand \bibitemNoStop [0]{.\EOS\space}%
\providecommand \EOS [0]{\spacefactor3000\relax}%
\providecommand \BibitemShut  [1]{\csname bibitem#1\endcsname}%
\let\auto@bib@innerbib\@empty
\bibitem [{\citenamefont {Kimble}(2008)}]{kimble2008quantum}%
  \BibitemOpen
  \bibfield  {author} {\bibinfo {author} {\bibfnamefont {H.~J.}\ \bibnamefont
  {Kimble}},\ }\href@noop {} {\bibfield  {journal} {\bibinfo  {journal}
  {Nature}\ }\textbf {\bibinfo {volume} {453}},\ \bibinfo {pages} {1023}
  (\bibinfo {year} {2008})}\BibitemShut {NoStop}%
\bibitem [{\citenamefont {Cirac}\ \emph {et~al.}(1997)\citenamefont {Cirac},
  \citenamefont {Zoller}, \citenamefont {Kimble},\ and\ \citenamefont
  {Mabuchi}}]{cirac1997quantum}%
  \BibitemOpen
  \bibfield  {author} {\bibinfo {author} {\bibfnamefont {J.~I.}\ \bibnamefont
  {Cirac}}, \bibinfo {author} {\bibfnamefont {P.}~\bibnamefont {Zoller}},
  \bibinfo {author} {\bibfnamefont {H.~J.}\ \bibnamefont {Kimble}}, \ and\
  \bibinfo {author} {\bibfnamefont {H.}~\bibnamefont {Mabuchi}},\ }\href@noop
  {} {\bibfield  {journal} {\bibinfo  {journal} {Phys. Rev. Lett.}\ }\textbf
  {\bibinfo {volume} {78}},\ \bibinfo {pages} {3221} (\bibinfo {year}
  {1997})}\BibitemShut {NoStop}%
\bibitem [{\citenamefont {Cirac}\ \emph {et~al.}(1999)\citenamefont {Cirac},
  \citenamefont {Ekert}, \citenamefont {Huelga},\ and\ \citenamefont
  {Macchiavello}}]{cirac1999distributed}%
  \BibitemOpen
  \bibfield  {author} {\bibinfo {author} {\bibfnamefont {J.~I.}\ \bibnamefont
  {Cirac}}, \bibinfo {author} {\bibfnamefont {A.~K.}\ \bibnamefont {Ekert}},
  \bibinfo {author} {\bibfnamefont {S.~F.}\ \bibnamefont {Huelga}}, \ and\
  \bibinfo {author} {\bibfnamefont {C.}~\bibnamefont {Macchiavello}},\ }\href
  {\doibase 10.1103/PhysRevA.59.4249} {\bibfield  {journal} {\bibinfo
  {journal} {Phys. Rev. A}\ }\textbf {\bibinfo {volume} {59}},\ \bibinfo
  {pages} {4249} (\bibinfo {year} {1999})}\BibitemShut {NoStop}%
\bibitem [{\citenamefont {Nakamura}\ \emph {et~al.}(1999)\citenamefont
  {Nakamura}, \citenamefont {Pashkin},\ and\ \citenamefont
  {Tsai}}]{nakamura1999coherent}%
  \BibitemOpen
  \bibfield  {author} {\bibinfo {author} {\bibfnamefont {Y.}~\bibnamefont
  {Nakamura}}, \bibinfo {author} {\bibfnamefont {Y.~A.}\ \bibnamefont
  {Pashkin}}, \ and\ \bibinfo {author} {\bibfnamefont {J.~S.}\ \bibnamefont
  {Tsai}},\ }\href@noop {} {\bibfield  {journal} {\bibinfo  {journal} {Nature}\
  }\textbf {\bibinfo {volume} {398}},\ \bibinfo {pages} {786} (\bibinfo {year}
  {1999})}\BibitemShut {NoStop}%
\bibitem [{\citenamefont {Devoret}\ and\ \citenamefont
  {Schoelkopf}(2013)}]{devoret2013superconducting}%
  \BibitemOpen
  \bibfield  {author} {\bibinfo {author} {\bibfnamefont {M.~H.}\ \bibnamefont
  {Devoret}}\ and\ \bibinfo {author} {\bibfnamefont {R.~J.}\ \bibnamefont
  {Schoelkopf}},\ }\href@noop {} {\bibfield  {journal} {\bibinfo  {journal}
  {Science}\ }\textbf {\bibinfo {volume} {339}},\ \bibinfo {pages} {1169}
  (\bibinfo {year} {2013})}\BibitemShut {NoStop}%
\bibitem [{\citenamefont {Strekalov}\ \emph {et~al.}(2009)\citenamefont
  {Strekalov}, \citenamefont {Schwefel}, \citenamefont {Savchenkov},
  \citenamefont {Matsko}, \citenamefont {Wang},\ and\ \citenamefont
  {Yu}}]{strekalov2009microwave}%
  \BibitemOpen
  \bibfield  {author} {\bibinfo {author} {\bibfnamefont {D.~V.}\ \bibnamefont
  {Strekalov}}, \bibinfo {author} {\bibfnamefont {H.~G.~L.}\ \bibnamefont
  {Schwefel}}, \bibinfo {author} {\bibfnamefont {A.~A.}\ \bibnamefont
  {Savchenkov}}, \bibinfo {author} {\bibfnamefont {A.~B.}\ \bibnamefont
  {Matsko}}, \bibinfo {author} {\bibfnamefont {L.~J.}\ \bibnamefont {Wang}}, \
  and\ \bibinfo {author} {\bibfnamefont {N.}~\bibnamefont {Yu}},\ }\href
  {\doibase 10.1103/PhysRevA.80.033810} {\bibfield  {journal} {\bibinfo
  {journal} {Phys. Rev. A}\ }\textbf {\bibinfo {volume} {80}},\ \bibinfo
  {pages} {033810} (\bibinfo {year} {2009})}\BibitemShut {NoStop}%
\bibitem [{\citenamefont {Rueda}\ \emph {et~al.}(2016)\citenamefont {Rueda},
  \citenamefont {Sedlmeir}, \citenamefont {Collodo}, \citenamefont {Vogl},
  \citenamefont {Stiller}, \citenamefont {Schunk}, \citenamefont {Strekalov},
  \citenamefont {Marquardt}, \citenamefont {Fink}, \citenamefont {Painter}
  \emph {et~al.}}]{rueda2016efficient}%
  \BibitemOpen
  \bibfield  {author} {\bibinfo {author} {\bibfnamefont {A.}~\bibnamefont
  {Rueda}}, \bibinfo {author} {\bibfnamefont {F.}~\bibnamefont {Sedlmeir}},
  \bibinfo {author} {\bibfnamefont {M.~C.}\ \bibnamefont {Collodo}}, \bibinfo
  {author} {\bibfnamefont {U.}~\bibnamefont {Vogl}}, \bibinfo {author}
  {\bibfnamefont {B.}~\bibnamefont {Stiller}}, \bibinfo {author} {\bibfnamefont
  {G.}~\bibnamefont {Schunk}}, \bibinfo {author} {\bibfnamefont {D.~V.}\
  \bibnamefont {Strekalov}}, \bibinfo {author} {\bibfnamefont {C.}~\bibnamefont
  {Marquardt}}, \bibinfo {author} {\bibfnamefont {J.~M.}\ \bibnamefont {Fink}},
  \bibinfo {author} {\bibfnamefont {O.}~\bibnamefont {Painter}},  \emph
  {et~al.},\ }\href@noop {} {\bibfield  {journal} {\bibinfo  {journal}
  {Optica}\ }\textbf {\bibinfo {volume} {3}},\ \bibinfo {pages} {597} (\bibinfo
  {year} {2016})}\BibitemShut {NoStop}%
\bibitem [{\citenamefont {Hafezi}\ \emph {et~al.}(2012)\citenamefont {Hafezi},
  \citenamefont {Kim}, \citenamefont {Rolston}, \citenamefont {Orozco},
  \citenamefont {Lev},\ and\ \citenamefont {Taylor}}]{hafezi2012atomic}%
  \BibitemOpen
  \bibfield  {author} {\bibinfo {author} {\bibfnamefont {M.}~\bibnamefont
  {Hafezi}}, \bibinfo {author} {\bibfnamefont {Z.}~\bibnamefont {Kim}},
  \bibinfo {author} {\bibfnamefont {S.~L.}\ \bibnamefont {Rolston}}, \bibinfo
  {author} {\bibfnamefont {L.~A.}\ \bibnamefont {Orozco}}, \bibinfo {author}
  {\bibfnamefont {B.~L.}\ \bibnamefont {Lev}}, \ and\ \bibinfo {author}
  {\bibfnamefont {J.~M.}\ \bibnamefont {Taylor}},\ }\href {\doibase
  10.1103/PhysRevA.85.020302} {\bibfield  {journal} {\bibinfo  {journal} {Phys.
  Rev. A}\ }\textbf {\bibinfo {volume} {85}},\ \bibinfo {pages} {020302}
  (\bibinfo {year} {2012})}\BibitemShut {NoStop}%
\bibitem [{\citenamefont {Petrosyan}\ \emph {et~al.}(2009)\citenamefont
  {Petrosyan}, \citenamefont {Bensky}, \citenamefont {Kurizki}, \citenamefont
  {Mazets}, \citenamefont {Majer},\ and\ \citenamefont
  {Schmiedmayer}}]{petrosyan2009reversible}%
  \BibitemOpen
  \bibfield  {author} {\bibinfo {author} {\bibfnamefont {D.}~\bibnamefont
  {Petrosyan}}, \bibinfo {author} {\bibfnamefont {G.}~\bibnamefont {Bensky}},
  \bibinfo {author} {\bibfnamefont {G.}~\bibnamefont {Kurizki}}, \bibinfo
  {author} {\bibfnamefont {I.}~\bibnamefont {Mazets}}, \bibinfo {author}
  {\bibfnamefont {J.}~\bibnamefont {Majer}}, \ and\ \bibinfo {author}
  {\bibfnamefont {J.}~\bibnamefont {Schmiedmayer}},\ }\href@noop {} {\bibfield
  {journal} {\bibinfo  {journal} {Phys. Rev. A}\ }\textbf {\bibinfo {volume}
  {79}},\ \bibinfo {pages} {040304} (\bibinfo {year} {2009})}\BibitemShut
  {NoStop}%
\bibitem [{\citenamefont {Williamson}\ \emph {et~al.}(2014)\citenamefont
  {Williamson}, \citenamefont {Chen},\ and\ \citenamefont
  {Longdell}}]{williamson2014magneto}%
  \BibitemOpen
  \bibfield  {author} {\bibinfo {author} {\bibfnamefont {L.~A.}\ \bibnamefont
  {Williamson}}, \bibinfo {author} {\bibfnamefont {Y.-H.}\ \bibnamefont
  {Chen}}, \ and\ \bibinfo {author} {\bibfnamefont {J.~J.}\ \bibnamefont
  {Longdell}},\ }\href@noop {} {\bibfield  {journal} {\bibinfo  {journal}
  {Phys. Rev. Lett.}\ }\textbf {\bibinfo {volume} {113}},\ \bibinfo {pages}
  {203601} (\bibinfo {year} {2014})}\BibitemShut {NoStop}%
\bibitem [{\citenamefont {Fernandez-Gonzalvo}\ \emph
  {et~al.}(2015)\citenamefont {Fernandez-Gonzalvo}, \citenamefont {Chen},
  \citenamefont {Yin}, \citenamefont {Rogge},\ and\ \citenamefont
  {Longdell}}]{fernandez2015coherent}%
  \BibitemOpen
  \bibfield  {author} {\bibinfo {author} {\bibfnamefont {X.}~\bibnamefont
  {Fernandez-Gonzalvo}}, \bibinfo {author} {\bibfnamefont {Y.-H.}\ \bibnamefont
  {Chen}}, \bibinfo {author} {\bibfnamefont {C.}~\bibnamefont {Yin}}, \bibinfo
  {author} {\bibfnamefont {S.}~\bibnamefont {Rogge}}, \ and\ \bibinfo {author}
  {\bibfnamefont {J.~J.}\ \bibnamefont {Longdell}},\ }\href@noop {} {\bibfield
  {journal} {\bibinfo  {journal} {Phys. Rev. A}\ }\textbf {\bibinfo {volume}
  {92}},\ \bibinfo {pages} {062313} (\bibinfo {year} {2015})}\BibitemShut
  {NoStop}%
\bibitem [{\citenamefont {Imamo{\u{g}}lu}(2009)}]{imamouglu2009cavity}%
  \BibitemOpen
  \bibfield  {author} {\bibinfo {author} {\bibfnamefont {A.}~\bibnamefont
  {Imamo{\u{g}}lu}},\ }\href@noop {} {\bibfield  {journal} {\bibinfo  {journal}
  {Phys. Rev. Lett.}\ }\textbf {\bibinfo {volume} {102}},\ \bibinfo {pages}
  {083602} (\bibinfo {year} {2009})}\BibitemShut {NoStop}%
\bibitem [{\citenamefont {Blum}\ \emph {et~al.}(2015)\citenamefont {Blum},
  \citenamefont {O'Brien}, \citenamefont {Lauk}, \citenamefont {Bushev},
  \citenamefont {Fleischhauer},\ and\ \citenamefont
  {Morigi}}]{blum2015interfacing}%
  \BibitemOpen
  \bibfield  {author} {\bibinfo {author} {\bibfnamefont {S.}~\bibnamefont
  {Blum}}, \bibinfo {author} {\bibfnamefont {C.}~\bibnamefont {O'Brien}},
  \bibinfo {author} {\bibfnamefont {N.}~\bibnamefont {Lauk}}, \bibinfo {author}
  {\bibfnamefont {P.}~\bibnamefont {Bushev}}, \bibinfo {author} {\bibfnamefont
  {M.}~\bibnamefont {Fleischhauer}}, \ and\ \bibinfo {author} {\bibfnamefont
  {G.}~\bibnamefont {Morigi}},\ }\href@noop {} {\bibfield  {journal} {\bibinfo
  {journal} {Phys. Rev. A}\ }\textbf {\bibinfo {volume} {91}},\ \bibinfo
  {pages} {033834} (\bibinfo {year} {2015})}\BibitemShut {NoStop}%
\bibitem [{\citenamefont {Hisatomi}\ \emph {et~al.}(2016)\citenamefont
  {Hisatomi}, \citenamefont {Osada}, \citenamefont {Tabuchi}, \citenamefont
  {Ishikawa}, \citenamefont {Noguchi}, \citenamefont {Yamazaki}, \citenamefont
  {Usami},\ and\ \citenamefont {Nakamura}}]{hisatomi2016bidirectional}%
  \BibitemOpen
  \bibfield  {author} {\bibinfo {author} {\bibfnamefont {R.}~\bibnamefont
  {Hisatomi}}, \bibinfo {author} {\bibfnamefont {A.}~\bibnamefont {Osada}},
  \bibinfo {author} {\bibfnamefont {Y.}~\bibnamefont {Tabuchi}}, \bibinfo
  {author} {\bibfnamefont {T.}~\bibnamefont {Ishikawa}}, \bibinfo {author}
  {\bibfnamefont {A.}~\bibnamefont {Noguchi}}, \bibinfo {author} {\bibfnamefont
  {R.}~\bibnamefont {Yamazaki}}, \bibinfo {author} {\bibfnamefont
  {K.}~\bibnamefont {Usami}}, \ and\ \bibinfo {author} {\bibfnamefont
  {Y.}~\bibnamefont {Nakamura}},\ }\href@noop {} {\bibfield  {journal}
  {\bibinfo  {journal} {Phys. Rev. B}\ }\textbf {\bibinfo {volume} {93}},\
  \bibinfo {pages} {174427} (\bibinfo {year} {2016})}\BibitemShut {NoStop}%
\bibitem [{\citenamefont {Das}\ \emph {et~al.}(2017)\citenamefont {Das},
  \citenamefont {Elfving}, \citenamefont {Faez},\ and\ \citenamefont
  {S{\o}rensen}}]{das2017interfacing}%
  \BibitemOpen
  \bibfield  {author} {\bibinfo {author} {\bibfnamefont {S.}~\bibnamefont
  {Das}}, \bibinfo {author} {\bibfnamefont {V.~E.}\ \bibnamefont {Elfving}},
  \bibinfo {author} {\bibfnamefont {S.}~\bibnamefont {Faez}}, \ and\ \bibinfo
  {author} {\bibfnamefont {A.~S.}\ \bibnamefont {S{\o}rensen}},\ }\href@noop {}
  {\bibfield  {journal} {\bibinfo  {journal} {Phys. Rev. Lett.}\ }\textbf
  {\bibinfo {volume} {118}},\ \bibinfo {pages} {140501} (\bibinfo {year}
  {2017})}\BibitemShut {NoStop}%
\bibitem [{\citenamefont {Tian}(2015)}]{tian2015optoelectromechanical}%
  \BibitemOpen
  \bibfield  {author} {\bibinfo {author} {\bibfnamefont {L.}~\bibnamefont
  {Tian}},\ }\href@noop {} {\bibfield  {journal} {\bibinfo  {journal} {Ann.
  Phys.}\ }\textbf {\bibinfo {volume} {527}},\ \bibinfo {pages} {1} (\bibinfo
  {year} {2015})}\BibitemShut {NoStop}%
\bibitem [{\citenamefont {Bochmann}\ \emph {et~al.}(2013)\citenamefont
  {Bochmann}, \citenamefont {Vainsencher}, \citenamefont {Awschalom},\ and\
  \citenamefont {Cleland}}]{bochmann2013nanomechanical}%
  \BibitemOpen
  \bibfield  {author} {\bibinfo {author} {\bibfnamefont {J.}~\bibnamefont
  {Bochmann}}, \bibinfo {author} {\bibfnamefont {A.}~\bibnamefont
  {Vainsencher}}, \bibinfo {author} {\bibfnamefont {D.~D.}\ \bibnamefont
  {Awschalom}}, \ and\ \bibinfo {author} {\bibfnamefont {A.~N.}\ \bibnamefont
  {Cleland}},\ }\href@noop {} {\bibfield  {journal} {\bibinfo  {journal} {Nat.
  Phys.}\ }\textbf {\bibinfo {volume} {9}},\ \bibinfo {pages} {712} (\bibinfo
  {year} {2013})}\BibitemShut {NoStop}%
\bibitem [{\citenamefont {Balram}\ \emph {et~al.}(2016)\citenamefont {Balram},
  \citenamefont {Davan{\c{c}}o}, \citenamefont {Song},\ and\ \citenamefont
  {Srinivasan}}]{balram2016coherent}%
  \BibitemOpen
  \bibfield  {author} {\bibinfo {author} {\bibfnamefont {K.~C.}\ \bibnamefont
  {Balram}}, \bibinfo {author} {\bibfnamefont {M.~I.}\ \bibnamefont
  {Davan{\c{c}}o}}, \bibinfo {author} {\bibfnamefont {J.~D.}\ \bibnamefont
  {Song}}, \ and\ \bibinfo {author} {\bibfnamefont {K.}~\bibnamefont
  {Srinivasan}},\ }\href@noop {} {\bibfield  {journal} {\bibinfo  {journal}
  {Nat. photonics}\ }\textbf {\bibinfo {volume} {10}},\ \bibinfo {pages} {346}
  (\bibinfo {year} {2016})}\BibitemShut {NoStop}%
\bibitem [{\citenamefont {Bagci}\ \emph {et~al.}(2014)\citenamefont {Bagci},
  \citenamefont {Simonsen}, \citenamefont {Schmid}, \citenamefont {Villanueva},
  \citenamefont {Zeuthen}, \citenamefont {Appel}, \citenamefont {Taylor},
  \citenamefont {S{\o}rensen}, \citenamefont {Usami}, \citenamefont
  {Schliesser} \emph {et~al.}}]{bagci2014optical}%
  \BibitemOpen
  \bibfield  {author} {\bibinfo {author} {\bibfnamefont {T.}~\bibnamefont
  {Bagci}}, \bibinfo {author} {\bibfnamefont {A.}~\bibnamefont {Simonsen}},
  \bibinfo {author} {\bibfnamefont {S.}~\bibnamefont {Schmid}}, \bibinfo
  {author} {\bibfnamefont {L.~G.}\ \bibnamefont {Villanueva}}, \bibinfo
  {author} {\bibfnamefont {E.}~\bibnamefont {Zeuthen}}, \bibinfo {author}
  {\bibfnamefont {J.}~\bibnamefont {Appel}}, \bibinfo {author} {\bibfnamefont
  {J.~M.}\ \bibnamefont {Taylor}}, \bibinfo {author} {\bibfnamefont
  {A.}~\bibnamefont {S{\o}rensen}}, \bibinfo {author} {\bibfnamefont
  {K.}~\bibnamefont {Usami}}, \bibinfo {author} {\bibfnamefont
  {A.}~\bibnamefont {Schliesser}},  \emph {et~al.},\ }\href@noop {} {\bibfield
  {journal} {\bibinfo  {journal} {Nature}\ }\textbf {\bibinfo {volume} {507}},\
  \bibinfo {pages} {81} (\bibinfo {year} {2014})}\BibitemShut {NoStop}%
\bibitem [{\citenamefont {Andrews}\ \emph {et~al.}(2014)\citenamefont
  {Andrews}, \citenamefont {Peterson}, \citenamefont {Purdy}, \citenamefont
  {Cicak}, \citenamefont {Simmonds}, \citenamefont {Regal},\ and\ \citenamefont
  {Lehnert}}]{andrews2014bidirectional}%
  \BibitemOpen
  \bibfield  {author} {\bibinfo {author} {\bibfnamefont {R.~W.}\ \bibnamefont
  {Andrews}}, \bibinfo {author} {\bibfnamefont {R.~W.}\ \bibnamefont
  {Peterson}}, \bibinfo {author} {\bibfnamefont {T.~P.}\ \bibnamefont {Purdy}},
  \bibinfo {author} {\bibfnamefont {K.}~\bibnamefont {Cicak}}, \bibinfo
  {author} {\bibfnamefont {R.~W.}\ \bibnamefont {Simmonds}}, \bibinfo {author}
  {\bibfnamefont {C.~A.}\ \bibnamefont {Regal}}, \ and\ \bibinfo {author}
  {\bibfnamefont {K.~W.}\ \bibnamefont {Lehnert}},\ }\href@noop {} {\bibfield
  {journal} {\bibinfo  {journal} {Nat. Phys.}\ }\textbf {\bibinfo {volume}
  {10}},\ \bibinfo {pages} {321} (\bibinfo {year} {2014})}\BibitemShut
  {NoStop}%
\bibitem [{\citenamefont {Delley}\ \emph {et~al.}()\citenamefont {Delley},
  \citenamefont {Kroner}, \citenamefont {F\"alt}, \citenamefont {Wegscheider},\
  and\ \citenamefont {Imamo\u{g}lu}}]{Delley2017}%
  \BibitemOpen
  \bibfield  {author} {\bibinfo {author} {\bibfnamefont {Y.}~\bibnamefont
  {Delley}}, \bibinfo {author} {\bibfnamefont {M.}~\bibnamefont {Kroner}},
  \bibinfo {author} {\bibfnamefont {S.}~\bibnamefont {F\"alt}}, \bibinfo
  {author} {\bibfnamefont {W.}~\bibnamefont {Wegscheider}}, \ and\ \bibinfo
  {author} {\bibfnamefont {A.}~\bibnamefont {Imamo\u{g}lu}},\ }\href@noop {}
  {\bibinfo  {journal} {arXiv:1704.01033}\ }\BibitemShut {NoStop}%
\bibitem [{\citenamefont {Frey}\ \emph {et~al.}(2012)\citenamefont {Frey},
  \citenamefont {Leek}, \citenamefont {Beck}, \citenamefont {Blais},
  \citenamefont {Ihn}, \citenamefont {Ensslin},\ and\ \citenamefont
  {Wallraff}}]{Frey2012}%
  \BibitemOpen
\bibfield  {journal} {  }\bibfield  {author} {\bibinfo {author} {\bibfnamefont
  {T.}~\bibnamefont {Frey}}, \bibinfo {author} {\bibfnamefont {P.}~\bibnamefont
  {Leek}}, \bibinfo {author} {\bibfnamefont {M.}~\bibnamefont {Beck}}, \bibinfo
  {author} {\bibfnamefont {A.}~\bibnamefont {Blais}}, \bibinfo {author}
  {\bibfnamefont {T.}~\bibnamefont {Ihn}}, \bibinfo {author} {\bibfnamefont
  {K.}~\bibnamefont {Ensslin}}, \ and\ \bibinfo {author} {\bibfnamefont
  {A.}~\bibnamefont {Wallraff}},\ }\href@noop {} {\bibfield  {journal}
  {\bibinfo  {journal} {Phys. Rev. Lett.}\ }\textbf {\bibinfo {volume} {108}},\
  \bibinfo {pages} {046807} (\bibinfo {year} {2012})}\BibitemShut {NoStop}%
\bibitem [{\citenamefont {Stockklauser}\ \emph {et~al.}(2017)\citenamefont
  {Stockklauser}, \citenamefont {Scarlino}, \citenamefont {Koski},
  \citenamefont {Gasparinetti}, \citenamefont {Andersen}, \citenamefont
  {Reichl}, \citenamefont {Wegscheider}, \citenamefont {Ihn}, \citenamefont
  {Ensslin},\ and\ \citenamefont {Wallraff}}]{Stockklauser2017}%
  \BibitemOpen
  \bibfield  {author} {\bibinfo {author} {\bibfnamefont {A.}~\bibnamefont
  {Stockklauser}}, \bibinfo {author} {\bibfnamefont {P.}~\bibnamefont
  {Scarlino}}, \bibinfo {author} {\bibfnamefont {J.~V.}\ \bibnamefont {Koski}},
  \bibinfo {author} {\bibfnamefont {S.}~\bibnamefont {Gasparinetti}}, \bibinfo
  {author} {\bibfnamefont {C.~K.}\ \bibnamefont {Andersen}}, \bibinfo {author}
  {\bibfnamefont {C.}~\bibnamefont {Reichl}}, \bibinfo {author} {\bibfnamefont
  {W.}~\bibnamefont {Wegscheider}}, \bibinfo {author} {\bibfnamefont
  {T.}~\bibnamefont {Ihn}}, \bibinfo {author} {\bibfnamefont {K.}~\bibnamefont
  {Ensslin}}, \ and\ \bibinfo {author} {\bibfnamefont {A.}~\bibnamefont
  {Wallraff}},\ }\href {\doibase 10.1103/PhysRevX.7.011030} {\bibfield
  {journal} {\bibinfo  {journal} {Phys. Rev. X}\ }\textbf {\bibinfo {volume}
  {7}},\ \bibinfo {pages} {011030} (\bibinfo {year} {2017})}\BibitemShut
  {NoStop}%
\bibitem [{\citenamefont {Krenner}\ \emph {et~al.}(2005)\citenamefont
  {Krenner}, \citenamefont {Sabathil}, \citenamefont {Clark}, \citenamefont
  {Kress}, \citenamefont {Schuh}, \citenamefont {Bichler}, \citenamefont
  {Abstreiter},\ and\ \citenamefont {Finley}}]{krenner2005}%
  \BibitemOpen
  \bibfield  {author} {\bibinfo {author} {\bibfnamefont {H.~J.}\ \bibnamefont
  {Krenner}}, \bibinfo {author} {\bibfnamefont {M.}~\bibnamefont {Sabathil}},
  \bibinfo {author} {\bibfnamefont {E.~C.}\ \bibnamefont {Clark}}, \bibinfo
  {author} {\bibfnamefont {A.}~\bibnamefont {Kress}}, \bibinfo {author}
  {\bibfnamefont {D.}~\bibnamefont {Schuh}}, \bibinfo {author} {\bibfnamefont
  {M.}~\bibnamefont {Bichler}}, \bibinfo {author} {\bibfnamefont
  {G.}~\bibnamefont {Abstreiter}}, \ and\ \bibinfo {author} {\bibfnamefont
  {J.~J.}\ \bibnamefont {Finley}},\ }\href {\doibase
  10.1103/PhysRevLett.94.057402} {\bibfield  {journal} {\bibinfo  {journal}
  {Phys. Rev. Lett.}\ }\textbf {\bibinfo {volume} {94}},\ \bibinfo {pages}
  {057402} (\bibinfo {year} {2005})}\BibitemShut {NoStop}%
\bibitem [{\citenamefont {Delteil}\ \emph {et~al.}(2016)\citenamefont
  {Delteil}, \citenamefont {Sun}, \citenamefont {Gao}, \citenamefont {Togan},
  \citenamefont {F\"alt},\ and\ \citenamefont {Imamo\u{g}lu}}]{delteil2016}%
  \BibitemOpen
  \bibfield  {author} {\bibinfo {author} {\bibfnamefont {A.}~\bibnamefont
  {Delteil}}, \bibinfo {author} {\bibfnamefont {Z.}~\bibnamefont {Sun}},
  \bibinfo {author} {\bibfnamefont {W.}~\bibnamefont {Gao}}, \bibinfo {author}
  {\bibfnamefont {E.}~\bibnamefont {Togan}}, \bibinfo {author} {\bibfnamefont
  {S.}~\bibnamefont {F\"alt}}, \ and\ \bibinfo {author} {\bibfnamefont
  {A.}~\bibnamefont {Imamo\u{g}lu}},\ }\href@noop {} {\bibfield  {journal}
  {\bibinfo  {journal} {Nat. Phys.}\ }\textbf {\bibinfo {volume} {12}},\
  \bibinfo {pages} {218} (\bibinfo {year} {2016})}\BibitemShut {NoStop}%
\bibitem [{\citenamefont {Delteil}\ \emph {et~al.}(2017)\citenamefont
  {Delteil}, \citenamefont {Sun}, \citenamefont {F\"alt},\ and\ \citenamefont
  {Imamo\u{g}lu}}]{delteil2017}%
  \BibitemOpen
  \bibfield  {author} {\bibinfo {author} {\bibfnamefont {A.}~\bibnamefont
  {Delteil}}, \bibinfo {author} {\bibfnamefont {Z.}~\bibnamefont {Sun}},
  \bibinfo {author} {\bibfnamefont {S.}~\bibnamefont {F\"alt}}, \ and\ \bibinfo
  {author} {\bibfnamefont {A.}~\bibnamefont {Imamo\u{g}lu}},\ }\href@noop {}
  {\bibfield  {journal} {\bibinfo  {journal} {Phys. Rev. Lett.}\ }\textbf
  {\bibinfo {volume} {118}},\ \bibinfo {pages} {177401} (\bibinfo {year}
  {2017})}\BibitemShut {NoStop}%
\bibitem [{\citenamefont {Carmichael}(1993)}]{carmichael1993quantum}%
  \BibitemOpen
  \bibfield  {author} {\bibinfo {author} {\bibfnamefont {H.~J.}\ \bibnamefont
  {Carmichael}},\ }\href@noop {} {\bibfield  {journal} {\bibinfo  {journal}
  {Phys. Rev. Lett.}\ }\textbf {\bibinfo {volume} {70}},\ \bibinfo {pages}
  {2273} (\bibinfo {year} {1993})}\BibitemShut {NoStop}%
\bibitem [{\citenamefont {Gardiner}(1993)}]{Gardiner1993}%
  \BibitemOpen
  \bibfield  {author} {\bibinfo {author} {\bibfnamefont {C.~W.}\ \bibnamefont
  {Gardiner}},\ }\href@noop {} {\bibfield  {journal} {\bibinfo  {journal}
  {Phys. Rev. Lett.}\ }\textbf {\bibinfo {volume} {70}},\ \bibinfo {pages}
  {2269} (\bibinfo {year} {1993})}\BibitemShut {NoStop}%
\bibitem [{\citenamefont {Pinotsi}\ and\ \citenamefont
  {Imamoglu}(2008)}]{pinotsi2008single}%
  \BibitemOpen
  \bibfield  {author} {\bibinfo {author} {\bibfnamefont {D.}~\bibnamefont
  {Pinotsi}}\ and\ \bibinfo {author} {\bibfnamefont {A.}~\bibnamefont
  {Imamoglu}},\ }\href@noop {} {\bibfield  {journal} {\bibinfo  {journal}
  {Phys. Rev. Lett.}\ }\textbf {\bibinfo {volume} {100}},\ \bibinfo {pages}
  {093603} (\bibinfo {year} {2008})}\BibitemShut {NoStop}%
\bibitem [{\citenamefont {Suri}\ \emph {et~al.}(2015)\citenamefont {Suri},
  \citenamefont {Keane}, \citenamefont {Bishop}, \citenamefont {Novikov},
  \citenamefont {Wellstood},\ and\ \citenamefont
  {Palmer}}]{PhysRevA.92.063801}%
  \BibitemOpen
  \bibfield  {author} {\bibinfo {author} {\bibfnamefont {B.}~\bibnamefont
  {Suri}}, \bibinfo {author} {\bibfnamefont {Z.~K.}\ \bibnamefont {Keane}},
  \bibinfo {author} {\bibfnamefont {L.~S.}\ \bibnamefont {Bishop}}, \bibinfo
  {author} {\bibfnamefont {S.}~\bibnamefont {Novikov}}, \bibinfo {author}
  {\bibfnamefont {F.~C.}\ \bibnamefont {Wellstood}}, \ and\ \bibinfo {author}
  {\bibfnamefont {B.~S.}\ \bibnamefont {Palmer}},\ }\href {\doibase
  10.1103/PhysRevA.92.063801} {\bibfield  {journal} {\bibinfo  {journal} {Phys.
  Rev. A}\ }\textbf {\bibinfo {volume} {92}},\ \bibinfo {pages} {063801}
  (\bibinfo {year} {2015})}\BibitemShut {NoStop}%
\bibitem [{\citenamefont {Peterer}\ \emph {et~al.}(2015)\citenamefont
  {Peterer}, \citenamefont {Bader}, \citenamefont {Jin}, \citenamefont {Yan},
  \citenamefont {Kamal}, \citenamefont {Gudmundsen}, \citenamefont {Leek},
  \citenamefont {Orlando}, \citenamefont {Oliver},\ and\ \citenamefont
  {Gustavsson}}]{peterer2015}%
  \BibitemOpen
  \bibfield  {author} {\bibinfo {author} {\bibfnamefont {M.~J.}\ \bibnamefont
  {Peterer}}, \bibinfo {author} {\bibfnamefont {S.~J.}\ \bibnamefont {Bader}},
  \bibinfo {author} {\bibfnamefont {X.}~\bibnamefont {Jin}}, \bibinfo {author}
  {\bibfnamefont {F.}~\bibnamefont {Yan}}, \bibinfo {author} {\bibfnamefont
  {A.}~\bibnamefont {Kamal}}, \bibinfo {author} {\bibfnamefont
  {T.}~\bibnamefont {Gudmundsen}}, \bibinfo {author} {\bibfnamefont {P.~J.}\
  \bibnamefont {Leek}}, \bibinfo {author} {\bibfnamefont {T.~P.}\ \bibnamefont
  {Orlando}}, \bibinfo {author} {\bibfnamefont {W.~D.}\ \bibnamefont {Oliver}},
  \ and\ \bibinfo {author} {\bibfnamefont {S.}~\bibnamefont {Gustavsson}},\
  }\href@noop {} {\bibfield  {journal} {\bibinfo  {journal} {Phys. Rev. Lett.}\
  }\textbf {\bibinfo {volume} {114}},\ \bibinfo {pages} {010501} (\bibinfo
  {year} {2015})}\BibitemShut {NoStop}%
\bibitem [{\citenamefont {Koch}\ \emph {et~al.}(2007)\citenamefont {Koch},
  \citenamefont {Yu}, \citenamefont {Gambetta}, \citenamefont {Houck},
  \citenamefont {Schuster}, \citenamefont {Majer}, \citenamefont {Blais},
  \citenamefont {Devoret}, \citenamefont {Girvin},\ and\ \citenamefont
  {Schoelkopf}}]{koch2007charge}%
  \BibitemOpen
  \bibfield  {author} {\bibinfo {author} {\bibfnamefont {J.}~\bibnamefont
  {Koch}}, \bibinfo {author} {\bibfnamefont {T.~M.}\ \bibnamefont {Yu}},
  \bibinfo {author} {\bibfnamefont {J.}~\bibnamefont {Gambetta}}, \bibinfo
  {author} {\bibfnamefont {A.~A.}\ \bibnamefont {Houck}}, \bibinfo {author}
  {\bibfnamefont {D.~I.}\ \bibnamefont {Schuster}}, \bibinfo {author}
  {\bibfnamefont {J.}~\bibnamefont {Majer}}, \bibinfo {author} {\bibfnamefont
  {A.}~\bibnamefont {Blais}}, \bibinfo {author} {\bibfnamefont {M.~H.}\
  \bibnamefont {Devoret}}, \bibinfo {author} {\bibfnamefont {S.~M.}\
  \bibnamefont {Girvin}}, \ and\ \bibinfo {author} {\bibfnamefont {R.~J.}\
  \bibnamefont {Schoelkopf}},\ }\href@noop {} {\bibfield  {journal} {\bibinfo
  {journal} {Phys. Rev. A}\ }\textbf {\bibinfo {volume} {76}},\ \bibinfo
  {pages} {042319} (\bibinfo {year} {2007})}\BibitemShut {NoStop}%
\bibitem [{\citenamefont {Chang}\ \emph {et~al.}(2013)\citenamefont {Chang},
  \citenamefont {Vissers}, \citenamefont {C{\'o}rcoles}, \citenamefont
  {Sandberg}, \citenamefont {Gao}, \citenamefont {Abraham}, \citenamefont
  {Chow}, \citenamefont {Gambetta}, \citenamefont {Rothwell}, \citenamefont
  {Keefe} \emph {et~al.}}]{chang2013improved}%
  \BibitemOpen
  \bibfield  {author} {\bibinfo {author} {\bibfnamefont {J.~B.}\ \bibnamefont
  {Chang}}, \bibinfo {author} {\bibfnamefont {M.~R.}\ \bibnamefont {Vissers}},
  \bibinfo {author} {\bibfnamefont {A.~D.}\ \bibnamefont {C{\'o}rcoles}},
  \bibinfo {author} {\bibfnamefont {M.}~\bibnamefont {Sandberg}}, \bibinfo
  {author} {\bibfnamefont {J.}~\bibnamefont {Gao}}, \bibinfo {author}
  {\bibfnamefont {D.~W.}\ \bibnamefont {Abraham}}, \bibinfo {author}
  {\bibfnamefont {J.~M.}\ \bibnamefont {Chow}}, \bibinfo {author}
  {\bibfnamefont {J.~M.}\ \bibnamefont {Gambetta}}, \bibinfo {author}
  {\bibfnamefont {M.~B.}\ \bibnamefont {Rothwell}}, \bibinfo {author}
  {\bibfnamefont {G.~A.}\ \bibnamefont {Keefe}},  \emph {et~al.},\ }\href@noop
  {} {\bibfield  {journal} {\bibinfo  {journal} {App. Phys. Lett.}\ }\textbf
  {\bibinfo {volume} {103}},\ \bibinfo {pages} {012602} (\bibinfo {year}
  {2013})}\BibitemShut {NoStop}%
\bibitem [{\citenamefont {Rigetti}\ \emph {et~al.}(2012)\citenamefont
  {Rigetti}, \citenamefont {Gambetta}, \citenamefont {Poletto}, \citenamefont
  {Plourde}, \citenamefont {Chow}, \citenamefont {C{\'o}rcoles}, \citenamefont
  {Smolin}, \citenamefont {Merkel}, \citenamefont {Rozen}, \citenamefont
  {Keefe} \emph {et~al.}}]{rigetti2012superconducting}%
  \BibitemOpen
  \bibfield  {author} {\bibinfo {author} {\bibfnamefont {C.}~\bibnamefont
  {Rigetti}}, \bibinfo {author} {\bibfnamefont {J.~M.}\ \bibnamefont
  {Gambetta}}, \bibinfo {author} {\bibfnamefont {S.}~\bibnamefont {Poletto}},
  \bibinfo {author} {\bibfnamefont {B.~L.~T.}\ \bibnamefont {Plourde}},
  \bibinfo {author} {\bibfnamefont {J.~M.}\ \bibnamefont {Chow}}, \bibinfo
  {author} {\bibfnamefont {A.~D.}\ \bibnamefont {C{\'o}rcoles}}, \bibinfo
  {author} {\bibfnamefont {J.~A.}\ \bibnamefont {Smolin}}, \bibinfo {author}
  {\bibfnamefont {S.~T.}\ \bibnamefont {Merkel}}, \bibinfo {author}
  {\bibfnamefont {J.~R.}\ \bibnamefont {Rozen}}, \bibinfo {author}
  {\bibfnamefont {G.~A.}\ \bibnamefont {Keefe}},  \emph {et~al.},\ }\href@noop
  {} {\bibfield  {journal} {\bibinfo  {journal} {Phys. Rev. B}\ }\textbf
  {\bibinfo {volume} {86}},\ \bibinfo {pages} {100506} (\bibinfo {year}
  {2012})}\BibitemShut {NoStop}%
\bibitem [{\citenamefont {Paik}\ \emph {et~al.}(2011)\citenamefont {Paik},
  \citenamefont {Schuster}, \citenamefont {Bishop}, \citenamefont {Kirchmair},
  \citenamefont {Catelani}, \citenamefont {Sears}, \citenamefont {Johnson},
  \citenamefont {Reagor}, \citenamefont {Frunzio}, \citenamefont {Glazman}
  \emph {et~al.}}]{paik2011observation}%
  \BibitemOpen
  \bibfield  {author} {\bibinfo {author} {\bibfnamefont {H.}~\bibnamefont
  {Paik}}, \bibinfo {author} {\bibfnamefont {D.~I.}\ \bibnamefont {Schuster}},
  \bibinfo {author} {\bibfnamefont {L.~S.}\ \bibnamefont {Bishop}}, \bibinfo
  {author} {\bibfnamefont {G.}~\bibnamefont {Kirchmair}}, \bibinfo {author}
  {\bibfnamefont {G.}~\bibnamefont {Catelani}}, \bibinfo {author}
  {\bibfnamefont {A.~P.}\ \bibnamefont {Sears}}, \bibinfo {author}
  {\bibfnamefont {B.~R.}\ \bibnamefont {Johnson}}, \bibinfo {author}
  {\bibfnamefont {M.~J.}\ \bibnamefont {Reagor}}, \bibinfo {author}
  {\bibfnamefont {L.}~\bibnamefont {Frunzio}}, \bibinfo {author} {\bibfnamefont
  {L.~I.}\ \bibnamefont {Glazman}},  \emph {et~al.},\ }\href@noop {} {\bibfield
   {journal} {\bibinfo  {journal} {Phys. Rev. Lett.}\ }\textbf {\bibinfo
  {volume} {107}},\ \bibinfo {pages} {240501} (\bibinfo {year}
  {2011})}\BibitemShut {NoStop}%
\bibitem [{\citenamefont {Barends}\ \emph {et~al.}(2013)\citenamefont
  {Barends}, \citenamefont {Kelly}, \citenamefont {Megrant}, \citenamefont
  {Sank}, \citenamefont {Jeffrey}, \citenamefont {Chen}, \citenamefont {Yin},
  \citenamefont {Chiaro}, \citenamefont {Mutus}, \citenamefont {Neill} \emph
  {et~al.}}]{barends2013coherent}%
  \BibitemOpen
  \bibfield  {author} {\bibinfo {author} {\bibfnamefont {R.}~\bibnamefont
  {Barends}}, \bibinfo {author} {\bibfnamefont {J.}~\bibnamefont {Kelly}},
  \bibinfo {author} {\bibfnamefont {A.}~\bibnamefont {Megrant}}, \bibinfo
  {author} {\bibfnamefont {D.}~\bibnamefont {Sank}}, \bibinfo {author}
  {\bibfnamefont {E.}~\bibnamefont {Jeffrey}}, \bibinfo {author} {\bibfnamefont
  {Y.}~\bibnamefont {Chen}}, \bibinfo {author} {\bibfnamefont {Y.}~\bibnamefont
  {Yin}}, \bibinfo {author} {\bibfnamefont {B.}~\bibnamefont {Chiaro}},
  \bibinfo {author} {\bibfnamefont {J.}~\bibnamefont {Mutus}}, \bibinfo
  {author} {\bibfnamefont {C.}~\bibnamefont {Neill}},  \emph {et~al.},\
  }\href@noop {} {\bibfield  {journal} {\bibinfo  {journal} {Phys. Rev. Lett.}\
  }\textbf {\bibinfo {volume} {111}},\ \bibinfo {pages} {080502} (\bibinfo
  {year} {2013})}\BibitemShut {NoStop}%
\bibitem [{\citenamefont {Dewes}\ \emph {et~al.}(2012)\citenamefont {Dewes},
  \citenamefont {Ong}, \citenamefont {Schmitt}, \citenamefont {Lauro},
  \citenamefont {Boulant}, \citenamefont {Bertet}, \citenamefont {Vion},\ and\
  \citenamefont {Esteve}}]{dewes2012}%
  \BibitemOpen
  \bibfield  {author} {\bibinfo {author} {\bibfnamefont {A.}~\bibnamefont
  {Dewes}}, \bibinfo {author} {\bibfnamefont {F.~R.}\ \bibnamefont {Ong}},
  \bibinfo {author} {\bibfnamefont {V.}~\bibnamefont {Schmitt}}, \bibinfo
  {author} {\bibfnamefont {R.}~\bibnamefont {Lauro}}, \bibinfo {author}
  {\bibfnamefont {N.}~\bibnamefont {Boulant}}, \bibinfo {author} {\bibfnamefont
  {P.}~\bibnamefont {Bertet}}, \bibinfo {author} {\bibfnamefont
  {D.}~\bibnamefont {Vion}}, \ and\ \bibinfo {author} {\bibfnamefont
  {D.}~\bibnamefont {Esteve}},\ }\href@noop {} {\bibfield  {journal} {\bibinfo
  {journal} {Phys. Rev. Lett.}\ }\textbf {\bibinfo {volume} {108}},\ \bibinfo
  {pages} {057002} (\bibinfo {year} {2012})}\BibitemShut {NoStop}%
\end{thebibliography}%

\end{document}


\title{SUPPLEMENTAL MATERIAL \\
Quantum interface between photonic and superconducting qubits}

\author{Yuta Tsuchimoto, Patrick Kn\"{u}ppel, Aymeric Delteil, Zhe Sun, Martin Kroner,}
\author{Ata\c{c} Imamo\u{g}lu}
\affiliation{Institute of Quantum Electronics, ETH Zurich, CH-8093 Z\"{u}rich, Switzerland}

\maketitle

\subsection{Quantum Monte Carlo method}

We consider optical to microwave conversion. In the quantum trajectory formalism, the time evolution of the system is given by the Schr\"odinger equation:
\begin{equation}
\frac{d}{dt}\ket{\psi(t)} = \frac{1}{i\hbar} H_\tx{eff} \ket{\psi(t)},
\end{equation}
where $\psi(t)$ is a stochastic wavefunction and $H_\tx{eff}$ is the non-Hermitian Hamiltonian given by
\begin{equation}
\label{hamiltonian}
H_\tx{eff} = H_\tx{s} + H_\tx{t} + H_\tx{st} - \frac{i \hbar}{2} \sum_k \hat{C}^\dagger_k \hat{C}_k.
\end{equation}
where $H_\tx{s}$ ($H_\tx{t}$) is the source (target) Hamiltonian and $H_\tx{st}$ the interaction Hamiltonian. The collapse operators $\hat{C}_k$ of this system are
\begin{equation}
\begin{aligned}
\hat{C}_1  &=  \sqrt{\Gamma_\tx{FE}^{(\tx{s})}} \sigma_\tx{EF}^{(\tx{s})} +  \sqrt{\Gamma_\tx{FG}^{(\tx{t})} \eta} \sigma_\tx{GF}^{(\tx{t})}, \\ 
\hat{C}_2  &= \sqrt{\Gamma_\tx{FG}^{(\tx{t})} (1-\eta)} \sigma_\tx{GF}^{(\tx{t})}, \\
\hat{C}_3  &=  \sqrt{\Gamma_\tx{EG}^{(\tx{t})}} \sigma_\tx{GE}^{(\tx{t})}, \\
\hat{C}_4  &=  \sqrt{\kappa_\tx{c}} \hat{a}_\tx{c},
\end{aligned}
\end{equation}
where $\sigma_{ij} = \ket{i}\bra{j}$ express the projection ($i=j$) and lowering or rising operator ($i \neq j$) respectively. (s) and (t) stand for source and target (interface). $\hat{C}_1$ corresponds to a detection event where photons are emitted either in the $\ket{F}_\tx{s} \rightarrow \ket{E}_\tx{s}$ or $\ket{F,0_\tx{c}}_\tx{t} \rightarrow \ket{G,0_\tx{c}}_\tx{t}$ transitions. The latter transition is coupled to the incident light with a coupling efficiency $\eta$. The operator $\hat{C}_2$ denotes an event originating from the $\ket{F,0_\tx{c}}_\tx{t} \rightarrow \ket{G,0_\tx{c}}_\tx{t}$ transition which does not couple to the incident mode. $\hat{C}_3$ describes an event associated with the $\ket{E,1_\tx{c}}_\tx{t} \rightarrow \ket{G,1_\tx{c}}_\tx{t}$ transition. Finally, $\hat{C}_4$ accounts for the cavity decay event. By substituting these collapse operators, equation~\ref{hamiltonian} becomes:
\begin{eqnarray}
H_\tx{eff} = && \hbar \Omega_\tx{L} (\sigma_\tx{GF}^{(\tx{s})} + \sigma_\tx{FG}^{(\tx{s})}) + \hbar g_\tx{c} (\hat{a}_\tx{c} \sigma_\tx{FE}^{(\tx{t})} + \hat{a}_\tx{c}^\dagger \sigma_\tx{EF}^{(\tx{t})})\nonumber\\ &&- i\hbar \frac{\Gamma_\tx{FE}^{(\tx{s})}}{2}\sigma_\tx{FF}^{(\tx{s})} - i\hbar \frac{\Gamma_\tx{FG}^{(\tx{t})}}{2}\sigma_\tx{FF}^{(\tx{t})} - i \hbar \frac{\Gamma_\tx{EG}^{(\tx{t})}}{2}\sigma_\tx{EE}^{(\tx{t})}\nonumber\\ && -i\hbar \sqrt{\Gamma_\tx{FG}^{(\tx{t})}\Gamma_\tx{FE}^{(\tx{s})}\eta} \sigma_\tx{EF}^{(\tx{s})} \sigma_\tx{FG}^{(\tx{t})} -i \hbar \frac{\kappa_c}{2}\hat{a}_\tx{c}^\dagger \hat{a}_\tx{c}.
\end{eqnarray}
Here, we neglected the term $\sigma_\tx{FE}^{(\tx{s})}\sigma_\tx{GF}^{(\tx{t})}$ describing the reverse process where an emitted photon from the $\ket{F,0_\tx{c}}_\tx{t} \rightarrow \ket{G,0_\tx{c}}_\tx{t}$ transition drives the $\ket{E}_\tx{s} \rightarrow \ket{F}_\tx{s}$ transition because we assume unidirectional coupling realized by a Faraday rotator or a chiral waveguide. We calculated the time evolution of the stochastic wave function using the quantum Monte Carlo wave function approach. We set the initial state as the ground states for both the source and interface and assumed that the cavity does not have microwave photons, i.e. $\ket{\psi_\tx{initial}} = \ket{G}_\tx{s}\ket{G,0_\tx{c}}_\tx{t}$. As the excitation laser pulse, we chose a Gaussian pulse  with a peak Rabi frequency $\Omega_0$ which is of the same order as $\Gamma_\tx{FG}^{(\tx{t})}$. The bandwidth of the pulse was set to be smaller than $\Gamma_\tx{FG}^{(\tx{t})}$. The generated single-photon pulse shape from the $\ket{F}_\tx{s} \rightarrow \ket{E}_\tx{s}$ transition was ensured to be Gaussian by keeping $\Omega_0/\Gamma_\tx{FE}^{(\tx{s})}$ small. $\eta$ was assumed to be 1.0. By assuming a realistic CQD and SC cavity, we set each parameter as follows: $\Gamma_\tx{FG}^{(\tx{t})}/2\pi = 300\us\tx{MHz}$, $\kappa_c/2\pi = 3 \us\tx{MHz}$, $g_c/2\pi = 50-400\us\tx{MHz}$, $\Gamma_\tx{EG}^{(\tx{t})}/2\pi = 0-750 \us \tx{MHz}$. Since the $\ket{E,1_\tx{c}}_\tx{t} \rightarrow \ket{G,1_\tx{c}}_\tx{t}$ transition heralds a successful optical-to-microwave photon conversion, we counted this event to estimate the conversion efficiency and rate.

Next, we consider microwave to optical conversion. Here, the collapse operators of this scheme are as follows:
\begin{equation}
\begin{aligned}
\hat{C'}_1 &= \sqrt{\Gamma_\tx{FG}^{(\tx{t})}} \sigma_\tx{GF}^{(\tx{t})}, \\
\hat{C'}_2 &= \sqrt{\Gamma_\tx{EG}^{(\tx{t})}} \sigma_\tx{GE}^{(\tx{t})}, \\
\hat{C'}_3 &= \sqrt{\kappa_\tx{c}} \hat{a}_\tx{c}.
\end{aligned}
\end{equation}
The effective Hamiltonian is then given by
\begin{eqnarray}
H_\tx{eff} =&& \hbar \Omega_{0} (\sigma_\tx{GE}^{(\tx{t})} + \sigma_\tx{EG}^{(\tx{t})}) + \hbar g_\tx{c} (\hat{a}_\tx{c} \sigma_\tx{FE}^{(\tx{t})} + \hat{a}_\tx{c}^\dagger \sigma_\tx{EF}^{(\tx{t})})\nonumber\\ && - i\hbar \frac{\Gamma_\tx{FG}^{(\tx{t})}}{2}\sigma_\tx{FF}^{(\tx{t})}
- i \hbar \frac{\Gamma_\tx{EG}^{(\tx{t})}}{2}\sigma_\tx{EE}^{(\tx{t})} -i \hbar \frac{\kappa_\tx{c}}{2} \hat{a}_\tx{c}^\dagger \hat{a}_\tx{c}
\end{eqnarray}
The initial state is $\ket{\psi_\tx{initial}} = \ket{G,1_\tx{c}}_\tx{t}$. The parameters used for this simulation are the same as those of the optical to microwave conversion except for $\Omega_0 = \Gamma_\tx{FG}^{(\tx{t})}/3$. We counted the decay event $\hat{C'}_1$ to estimate the conversion efficiency and rate.

\subsection{Analytical formula for the optical-to-microwave conversion efficiency}
We consider a simple case where weak coherent field resonantly couples the target CQD with perfect mode matching $\eta = 1.0$ based on van Enk \cite{vanEnk}. Here, the coherent field satisfies 
\begin{equation}
b_\tx{in}(t)\ket{\beta} = \beta \tx{exp}(-i\omega_\tx{in}t)\ket{\beta},
\end{equation}
where $b_\tx{in}$ is the input field operator, $\omega_\tx{in}$ is the incident photon frequency, and $\beta$ is the amplitude of the input field.
Based on the input-output formalism, we write the mean output field $\braket{b_\tx{out}}$ generated from the $\ket{F,0_\tx{c}}_\tx{t} \rightarrow \ket{G,0_\tx{c}}_\tx{t}$ transition as follows:
\begin{equation}
\braket{b_\tx{out}} = \beta+\sqrt{\Gamma_\tx{FG}^{(\tx{t})}}\braket{\sigma_\tx{FG}^{(\tx{t})}}.
\end{equation}
Assuming that $\kappa_\tx{c}$ is sufficiently smaller than $g_\tx{c}$ and all the other decay rates,
we find that an analytical formula of the mean field is
\begin{equation}
\braket{b_\tx{out}} = \beta-\frac{2\beta\Gamma_\tx{EG}^{(\tx{t})}}{4g_\tx{c}^2/\Gamma_\tx{FG}^{(\tx{t})}+\Gamma_\tx{EG}^{(\tx{t})}-ig_\tx{c}}.
\end{equation}
Here, we define a normalized mean field $\braket{b_\tx{out}^\tx{n}} = \braket{b_\tx{out}}/\beta$. The efficiency of the $\ket{F,0_\tx{c}}_\tx{t} \rightarrow \ket{G,0_\tx{c}}_\tx{t}$ transition is given by the conjugate product $\braket{b_\tx{out}^\tx{n\dagger} b_\tx{out}^\tx{n}}$.
One can calculate the conversion efficiency $\zeta$ as the complement of this decay efficiency as shown in the main text.

\begin{figure*}
  \begin{center}
    \includegraphics[width=0.75\textwidth]{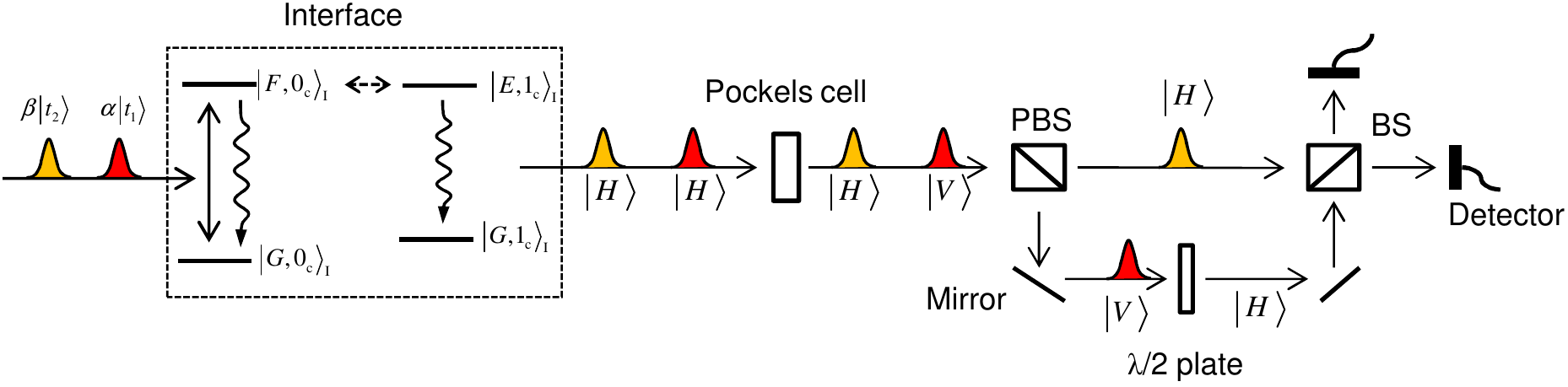}
  \end{center}
  \caption{Scheme of the optical setup proposed to erase the time-bin information from the herald photons. BS: beam splitter; PBS: polarized beam splitter. The red and yellow pulses show each time-bin component.}
  \label{S1}
\end{figure*}
\subsection{Erasure of time-bin information of the herald photons}

Figure \ref{S1} shows a proposed optical setup which compensates for the delay between the two time-bin components of photons emitted from the $\ket{E,1_\tx{c}}_\tx{t}\rightarrow \ket{G,0_\tx{c}}_\tx{t}$ transition, which allows to remove entanglement between the transferred qubits and the herald photons. The polarization of these photons is defined by the QD selection rules. Here, we note $\ket{H}$ ($\ket{V}$) the horizontally (vertically) polarized state, and we assume that the photons are initially $H$-polarized. A Pockels-cell driven synchronously with the state transfer protocol selectively rotates the early component from $\ket{H}$ to $\ket{V}$. This allows to use a polarized beam splitter to channel the two time components into two separate optical paths, such that the early component is delayed by $t_2 - t_1$ and rotated back to $\ket{H}$. Therefore the two components arrive simultaneously and in the same polarization state onto a second (non-polarized) beam splitter. A single-photon detector placed at one of the output port heralds a successful operation. This process is inherently probabilistic but does not require additional local operation on the qubit. On the other hand, the scheme can be in principle rendered deterministic by using two single-photon detectors of high ($\sim 100\%$) efficiency at the two output ports of the beam splitter. In this case, an additional qubit rotation would have to be performed on the transmon qubit state depending on which detector has clicked, due to the $\pi$ phase difference between the final qubit states left after a photon detection in either of the two output ports of the beam splitter.
\subsection{Coupling strength}
We show that the large dipole of the CQD interacting with enhanced cavity vacuum electric field leads to large $g_\tx{c}$ necessary for the fast and high-efficient conversion. The dipole moment of a CQD is given by $p \sim a \cdot e/2$ where $e$ is a charge of an electron and $a$ is the distance between the two quantum dot layers. We assume a typical distance $a \sim 10\us\text{nm}$, resulting in $p \sim 8 \times 10^{-28} \us\text{C$\cdot$m}$. This large dipole moment interacts with the cavity vacuum electric field given by
\begin{equation}
      E_\tx{rms} = \sqrt{\frac{\hbar\omega_\tx{c}}{2\epsilon_\tx{0}\epsilon_\tx{eff}V_\tx{eff}}}
   \label{eq:Vrms}
\end{equation}
where $\omega_\tx{c}$ is the resonant frequency of the SC cavity. $\epsilon_\tx{0}$ and $\epsilon_\tx{eff}$ are the permittivity of vacuum and the effective dielectric constant of the SC cavity, respectively.
This vacuum field can also be expressed as a function of the cavity impedance as follows:
\begin{equation}
   E_\tx{rms} = \sqrt{\frac{Z_\tx{cav}\hbar\omega_\tx{c}^2}{\pi d^2}}. 
   \label{eq:Evac}
\end{equation} 
where $d$ is the the gap between the center and ground conductor. $Z_\tx{cav}$ is the characteristic impedance of the SC cavity. For a typical SC cavity, $Z_\tx{cav}$ is about $50 \us{\Omega}$. A high impedance cavity can be used to enhance the vacuum field~\cite{Stoc17}. If we assume $Z_\tx{cav} \sim 2000 \us \Omega$, $d \sim 7 \us\mu\text{m}$ and $\omega_\tx{c}/2\pi \sim11\us\text{GHz}$, the vacuum field is  $E_\text{rms} \sim 3\us\text{V/m}$. Furthermore, $E_\text{rms}$ is enhanced between the top gate and the ground plane by factor of $\sqrt{\epsilon_\tx{eff}/\epsilon_\tx{GaAs}} \cdot d/d'$ (see Fig.~1 in the main text). $\epsilon_\tx{GaAs}$ is the dielectric constant of GaAs, which is about 13. $\epsilon_\tx{eff}$ is given by $(\pi c/l\omega_\tx{c})^2$ where $c$ is the speed of light in vacuum and $l$ is the length of the cavity. We assumed $l \sim3\us\text{mm}$ and $d' \sim 200 \us\text{nm}$ as realistic values. The coupling strength is therefore
\begin{equation}
  g_\tx{c}/2\pi = \frac{p \cdot \sqrt{\epsilon_\tx{eff}/\epsilon_\tx{GaAs}} \cdot d/d' \cdot E_\tx{rms}}{2\pi\hbar} \sim 200\us\text{MHz}.
  \label{eq:g}
\end{equation}
This large coupling strength satisfies the conditions for the high conversion efficiency and rate discussed in the paper.

{}